\pdfoutput=1
\documentclass[aps,prd,superscriptaddress,twocolumn,nofootinbib]{revtex4-1}


\usepackage[dvipsnames]{xcolor}
\usepackage{amsmath}
\usepackage{amssymb}
\usepackage{graphicx}
\usepackage{bm}
\usepackage{ulem} 
\usepackage{subcaption}
\usepackage{ragged2e}
\DeclareCaptionJustification{justified}{\justifying}
\usepackage{lipsum}

\usepackage[colorlinks=true,bookmarks=false,citecolor=blue,linkcolor=red,hyperfootnotes=true,urlcolor=blue]{hyperref}

\newcommand{\be}{\begin{equation}}
\newcommand{\ee}{\end{equation}}
\newcommand{\bea}{\begin{eqnarray}}
\newcommand{\eea}{\end{eqnarray}}

\newcommand{\p}{\partial}
\newcommand{\s}{\sigma}

\newcommand{\la}{\langle}
\newcommand{\ra}{\rangle}
\newcommand{\rd}{\mbox{d}}
\newcommand{\ri}{\mbox{i}}
\newcommand{\re}{\mbox{e}}

\renewcommand{\vec}[1]{{\bm #1}}

\begin{document}

\title{When cold, dense quarks in 1+1 and 3+1 dimensions are not a Fermi liquid}

\author{Marton Lajer}
\affiliation{ Division of  Condensed Matter Physics and Material Science,
Brookhaven National Laboratory, Upton, NY 11973-5000, USA}  

  \author{ Robert  M. Konik}                                                                                                 
\affiliation{ Division of  Condensed Matter Physics and Material Science,
Brookhaven National Laboratory, Upton, NY 11973-5000, USA}

\author{Robert D. Pisarski}
\affiliation{Department of Physics, Brookhaven National Laboratory, Upton, NY 11973}

\author{ Alexei  M. Tsvelik}                                                                                                 
\affiliation{ Division of  Condensed Matter Physics and Material Science,
Brookhaven National Laboratory, Upton, NY 11973-5000, USA}  

\begin{abstract}
We analyze the behavior of quarks coupled to a $SU(N_c)$ gauge theory in 1+1 dimensions.  In the limit of
strong coupling, the model reduces to a Wess-Zumino-Novikov-Witten (WZNW) model.
At nonzero density, excitations near the Fermi surface form a non-Fermi liquid.
With $N_f$ flavors, the finite density of quarks reduce to a
free $U(1)$ field, which governs fluctuations in baryon number, 
together with a WZNW $SU(N_f)$ nonlinear sigma model at level $N_c$, from the pion/kaon modes.
We compute the singularity in the
charge susceptibility at the Fermi surface and the attendant power law correlations.  We suggest that this
is relevant to the quarkyonic regime of cold, dense QCD in 3+1 dimensions,
in the limit that the Fermi surface is covered by many small patches, and the theory is effectively one dimensional.
In this regime the dominant excitations near the Fermi surface are not baryons, but gapless bosonic modes.
\end{abstract}

\maketitle

The behavior of QCD at nonzero temperature and density is of fundamental interest.  Since numerical simulations
on the lattice are not possible for three (or more) colors at nonzero density, and as quantum computers are far
from able to compute the behavior of cold, dense QCD,
it is useful to study analytically tractable models, even those in fewer dimensions, that may reflect the
essential features of the underlying physics.

At low density, nuclear matter is well described by effective theories of nucleons.  This describes the theory up to
and for some region beyond saturation density
\cite{Walecka:1995mi,Lalazissis:1996rd,Serot:1997xg,Akmal:1998cf,Carriere:2002bx,Danielewicz:2002pu,Epelbaum:2008ga,Machleidt:2011zz}.  Conversely, at very high density, perturbation theory in the QCD
coupling applies \cite{Kurkela:2009gj,Kurkela:2014vha,Kurkela:2016was,Ghisoiu:2016swa,Annala:2017llu,Gorda:2018gpy,Annala:2019puf,Gorda:2021znl,Gorda:2021kme}.
Previously it has been argued that at intermediate densities there is a quarkyonic regime, where the free
energy is that of an interacting theory of quarks and gluons, but near the Fermi surface, the excitations are confined
\cite{McLerran:2007qj,Andronic:2009gj,Kojo:2009ha,Kojo:2010fe,Kojo:2011cn,Fukushima:2015bda,McLerran:2018hbz,Jeong:2019lhv,Duarte:2020kvi,Duarte:2020xsp,Sen:2020peq,Sen:2020qcd,Zhao:2020dvu,Pisarski:2021aoz,Duarte:2021tsx}.
This is the regime that we wish to concentrate upon in this paper.  

It was shown in Refs.
\cite{Kojo:2010fe,Pisarski:2018bct,Pisarski:2020dnx,Pisarski:2021aoz,Pisarski:2021qof,Tsvelik:2021ccp}
that  the quarkyonic regime is a critical state which 
shares many aspects with systems in 1+1 dimensions.
We explore this analogy further in this paper.  

In 1+1 dimensions the gauge coupling $g$ has dimensions of mass, so weak coupling is the regime where
the quark mass is large relative to $g$.
We concentrate on strong coupling, where the quark masses are small relative to $g$.
For vanishing bare quark masses, non-Abelian bosonization demonstrates that the gapped color sector decouples.
In the vacuum, and at energies small relative to the gauge coupling,
the quarks are confined and the low energy degrees of freedom are gapless $U(1)$ and SU$(N_f)_{N_c}$
WZNW meson fields.  There are baryons, but these are not coherent excitations, but nonlocal
combinations of the mesonic fields. At small but finite quark mass,
the color sector can be integrated out, and generates relevant perturbations in the low energy sector.

In this paper we consider whether a non-Fermi liquid exists in QCD in 1+1 dimensions at nonzero density.
The Nambu-Jona-Lasinio (NJL) model was studied in 1+1 dimensions
in Ref. \cite{Azaria:2016mqb}.  At nonzero density, even at nonzero bare quark mass 
sectors with different symmetry decouple.
We argue that this  decoupling remains valid for QCD.
In the NJL model, there are two regimes at nonzero density:
a state with gapped flavor and gapless $U(1)$ excitations at low density, and a critical state at higher density.
The baryons are formed from the $U(1)$ and flavor fields, but in both regimes
the spectral function for a single baryon never exhibits a sharp peak, as the baryons
are ``incoherent''.
This is unlike the typical Fermi liquid, where the spectral function has a sharp peak, which allows for
the consistent definition of quasiparticles.
In the NJL model, the behavior of baryon Green's functions changes between the two regimes.
At low density, both the flavor sector and baryons are gapped,
and baryon correlation functions are exponentially damped.
In contrast, the high density phase is critical, as the baryon spectral function is gapless,
and baryon correlation functions fall off as a power law.  This can be termed a strange metal.

To establish the density at which a strange metal appears
depends upon the details of the low energy theory.
We argue that in 1+1 dimensions, the low energy theory is the same for both the NJL model and  QCD.
At sufficiently low density the quark mass term generates a relevant perturbation in the flavor sector.
For this to happen the Luttinger parameter in the
$U(1)$ sector must exceed a critical value determined by the quark mass,
the chemical potential, and how many flavors, $N_f$, and colors, $N_c$, there are.
We extract this dependence for two cases.
First, for an arbitrary $N_c$ and a single flavor, $N_f=1$
\cite{Baluni:1980bw,Steinhardt:1980ry,Cohen:1982mq,Gepner:1984au,Frishman:1992mr,Armoni:1998ny}.
Second, for two colors and two flavors, $N_c = N_f = 2$.  We expect that the behavior for
other values of $N_c$ and $N_f$ is qualitatively similar, although the details certainly change.
 
Our methods are only useful in the strong coupling regime of light quarks.
It is possible that for a low density of heavy quarks, 
QCD in 1+1 dimensions is no longer a Luttinger liquid, and baryon correlation functions are gapped,
like the low density regime of the NJL model \cite{Azaria:2016mqb}.
Heavy quarks inevitably form 
a Luttinger liquid at high density, though, once the mass term is negligible relative to the chemical potential.
It is also natural to conjecture that for any mass, baryons are incoherent at all densities.

A gauge theory in 1+1 dimensions coupled to fermions in the adjoint representation
has been studied in Refs. \cite{Gopakumar_2012} and \cite{Isachenkov_2014}.
For massless fields, at nonzero density
the theory is again a strange metal \cite{Gopakumar_2012},
described by a supersymmetric Wess-Zumino-Novikov-Witten (WZNW) model.  
This is considerably more complicated than the simple Luttinger liquid which emerges for
quarks in the fundamental representation.

While the phenomenology of QCD in 1+1 dimensions
is interesting in and of itself,
here we are primarily motivated by its consequences to quarkyonic matter in 3+1 dimensions.
For ordinary nuclear matter, the excitations
near the Fermi surface are baryons, and are typically gapped,
being Bogolyubov quasiparticles of either the superfluid or superconducting state.
While the gaps are small, on the order of a few MeV, transport coefficients are exponentially suppressed at temperatures
less than these gaps.
As the density increases, nuclear matter is quarkyonic, and forms a smectic liquid crystal
\cite{Kojo:2010fe,Pisarski:2018bct,Pisarski:2020dnx}.
The  Fermi surface of the theory spontaneously divides itself into patches, where the propagators for gapless bosons
are anisotropic, and differ
between the direction of the condensate and transverse fluctuations.
As was discussed previously \cite{Kojo:2010fe,Pisarski:2018bct,Pisarski:2020dnx},
the solution is stable when there are many  patches, when the net quark density is large.
Then the theory for each patch is effectively one dimensional, 
and the coherent excitations at low energy are then not baryons, but 
gapless bosons, both Abelian and non-Abelian, with a dispersion relation which is quasi-one-dimensional.
These bosonic modes are color singlets, related to a low-energy Wess-Zumino-Novikov-Witten (WZNW) model
of the flavor degrees of freedom.  Our arguments are qualitative, as our purpose is
simply to emphasize that the quarkyonic matter is a non-Fermi liquid.  This is of direct consequence
for the transport properties of cold quarykyonic matter.

In Sec. \ref{densematter} we analyze QCD, coupled to quarks in the fundamental representation,
in 1+1 dimensions.  The effective Wess-Zumino-Witten
Lagrangian at low energy is given in Sec. \ref{lowenergy22}, and extended to nonzero density in Sec. \ref{nonzerodensity}.
Detailed computations of the Luttinger parameter are given in Secs. \ref{Nf1TBA} and \ref{PTsubsec}, while the
properties of the strange metal for $N_c=3$, $N_f=2$ are detailed in Sec. \ref{Strangemetal32}.
In Sec. \ref{sec1d3d} we outline the possible relation
to cold, dense QCD in 3+1 dimensions.  Appendices include a review of the Quantum Ising model,
Appendix \ref{app_ising}, of the Thermodynamic Bethe Ansatz, Appendix  \ref{app_tba},
the numerical analysis for a single flavor, Appendix \ref{app_num},
and details of the perturbative calculation, Appendix \ref{app_details}.

\section{Dense matter for QCD in 1+1 dimensions}     \label{densematter}

 As we have discussed above, some physical aspects of the quarkyonic state in 3+1 dimensions
 are essentially  those of 1+1 dimensions. Therefore we will start our discussion from this case.

We will work with the premise that all physical states are color singlets.  Thus to obtain an effective low energy description, we have to integrate out the color degrees of freedom. The effective action and operators should be projected onto the color singlet ground state.
One way to approach the problem is to use non-Abelian bosonization and the notion of conformal embeddings.
Here we will use
the fact that the Hamiltonian of massless Dirac fermions of $N_c$
colors and $N_f$ flavors can be decomposed into three commuting Wess-Zumino-Novikov-Witten Hamiltonians describing the U(1), the color, and the flavor sectors
\cite{Baluni:1980bw,Steinhardt:1980ry,Cohen:1982mq,Gepner:1984au,Frishman:1992mr,Armoni:1998ny}.

In the NJL model there is a local interaction between color currents and the
color sector is gapped \cite{Azaria:2016mqb}.
In QCD, integrating out the gluons generates a long range interaction between the color currents; thus
it is reasonable to assume that the color sector remains gapped.
In 1+1 dimensions, the gap is of order $\Lambda_{QCD} \sim g$.  For massless quarks, the color currents commute
with the charge and flavor sectors, leaving a $U(N_f)_{N_c}$ WZNW model at low energies.

There are issues which need clarification related to the projection of the observables.
They emerge even for the case of
massless quarks or for the dense phase where the mass term becomes irrelevant. As was noted in the earlier works
on QCD in 1+1 dimensions,
there are various observables, including the mass term,  which do not
factorize into a product of local fields from the  color and flavor
sectors. This was noticed already  in Ref. \cite{Frishman:1992mr},
where it was shown that the Hamiltonian of QCD in 1+1 dimensions,
for massless quarks with $N_f,N_c \geq 2$, can
be decomposed via a conformal embedding into three commuting parts.
These three are WZNW models
acting respectively on the charge, the flavor and the color sectors.
Even so, the mass term cannot be written as a product of
mutually local fields of these models. In fact, this is a common
problem for conformal embeddings.
After integrating out the high energy degrees of freedom,
the color singlet operators can be expressed as local
fields of the low energy field theory, which is the $U(N_f)_{N_c}$ WZNW
model. Below we discuss the details of the projection
for two particular cases: a single flavor, $N_f=1$, for arbitrary $N_c$, and  $N_c = N_f =2$.
The conformal embedding is done using Abelian bosonization, as
in Ref. \cite{Controzzi_2005}. The conclusion is that after one
executes a projection onto the color singlet vacuum, the corresponding fields are
expressible in terms of the local fields of the charge and flavor sector.

The QCD Lagrangian with the quarks in the fundamental representation of the SU(N$_c$) color group is
\bea
&& L = \int \rd^D x\Big[-\frac{1}{8\pi g^2} {\rm tr} F^{\mu\nu}F_{\mu\nu} +\nonumber\\
&& \bar q_{f,\s}\gamma^{\mu}D_{\mu,\sigma\sigma^\prime}q_{f,\s'} + m\bar q_{f,\s}q_{f,\s}
\Big],
\eea
where the covariant derivative is defined as
\be
D_\mu^{\sigma\sigma^\prime}=\delta^{\sigma\sigma^\prime}\partial_\mu-i A_\mu^{\sigma\sigma^\prime}.
\ee
The quark fields $\bar q, q$ carry $f=1,...N_f$ flavor and $\s=1,...N_c$ color indices and the color Yang-Mills field is
\bea
F_{\mu\nu} = \p_{\mu}A_{\nu} - \p_{\nu}A_{\mu} -i [A_{\mu},A_{\nu}].
\eea
In 1+1 dimensions one can choose  the gauge where $A_x =0$ and integrate out
the color Yang-Mills field. Then after a suitable choice of the gamma matrices, we arrive at the following Hamiltonian
for QCD in 1+1 dimensions:
\bea
H &=& \sum_{f=1}^{N_f}\sum_{\s=1}^{N_c} \int \rd x \Big[-i R^+_{f,\s}\p_xR_{f,\s} + i L^+_{f,\s}\p_x L_{f,\s} \cr\cr
&-& m(R_{f,\s}^+L_{f,\s} +L^+_{f,\s}R_{f,\s})\Big]\cr\cr
&-& \pi g^2\int\rd x\rd y  J^a_0(x)|x-y|J_0^a(y), \label{QCD1}
\eea
where
\bea
J_0^a &=& J_R^a + J_L^a, \\
J^a_R &=& \sum_{f=1}^{N_f}R^+_{f,\s}t^a_{\s\s'}R_{f,\s'}, ~~ J^a_L = \sum_{f=1}^{N_f}L^+_{f,\s}t^a_{\s\s'}L_{f,\s'}\label{qcdH}.
\eea
Here  $t^a$-matrices are generators of the $su(N_c)$ algebra
and $R,L$ are the right- and left moving components of the $q$-spinor field.  The currents of the right- and left moving quarks $J_{R,L}$ obey a $SU(N_c)_{N_f}$ Kac-Moody algebra. (For a comprehensive review of non-Abelian bosonization, see Ref. \cite{James:2017cpc}.)
One can rewrite the Hamiltonian as the sum of three mutually commuting parts plus the mass term, which mixes them all:
\bea
H = H[U(1)] + H_{color}+ H_{flavor} + \mbox{mass},
\eea
where $H[U(1)], H_{flavor}$ are Hamiltonians of the Wess-Zumino-Novikov-Witten (WZNW) models of the corresponding groups.
$H_{color}$ is the $SU(N_c)_{N_f}$ WZNW model perturbed by the current-current interaction of Eq. (\ref{QCD1}).
These Hamiltonians are expressed in terms of the U(1), $SU(N_c)_{N_f}$, and $SU(N_f)_{N_c}$ Kac-Moody currents.

 For $N_f > 1$ at zero quark mass the Hamiltonian is separated into
 three commuting parts.  The mass term however violates this
 separation. As was mentioned in the Introduction, the corresponding
 density operator  cannot be even  factorized into a product of
 mutually local operators acting on the corresponding Hilbert
 spaces. This may lead to difficulties in formulating the low energy
 theory in the strong coupling limit. We argued in
 Ref. \cite{Azaria:2016mqb} for the NJL model, the issue is resolved
 once the Hamiltonian is projected on the color singlet ground state
 of the color sector. To elucidate the nature of the issues involved
 we will consider both the case $N_f=1$ and the case of $N_c = N_f
 =2$.  Here one can treat the conformal embedding using Abelian bosonization, as
 was done, for example, in Ref. \cite{Controzzi_2005}.

\subsection{Low energy theory for $N_f =1$}

We being with the simplest case, that of a single flavor, $N_f = 1$.  We start with Abelian bosonization of the free quarks and use the
 standard bosonization rules for the operators of right- and left moving quarks ($\s= 1,...N_c; ~~f=1,...N_f$):
\bea
R_{f,\s} &=& \frac{1}{\sqrt{2\pi}}\xi_{f,\s} \exp[i\sqrt{4\pi}\varphi_{f,\s}], \nonumber\\
L_{f,\s} &=& \frac{1}{\sqrt{2\pi}}\xi_{f,\s}\exp[-i\sqrt{4\pi}\bar\varphi_{f,\s}],
\eea
where $\varphi_n, \bar\varphi_n$ $(n = (f,\s)$) are chiral bosonic fields governed by the Gaussian action (from now on we will work in Euclidean time),
\bea
S_{Gauss} = \frac{1}{2}\sum_{n= (f,\s)}\int \rd\tau\rd x \left(\partial_\tau\varphi_n\right)^2+\left(\partial_x\varphi_n\right)^2.\label{Bose}
\eea
The chiral vertex operators, i.e. $\exp[i\sqrt{4\pi}\bar\varphi_{f,\s}]$, are assumed to be normal ordered and have a normalization determined by
\begin{equation}
\la\la e^{-i\sqrt{4\pi}\varphi_{f,\s}(\tau,x)} e^{i\sqrt{4\pi}\varphi_{f,\s}(0,0)}\ra\ra = \frac{1}{\tau -i x}.
\end{equation}
Finally $\xi_n$ are anticommuting Klein factors that ensure this bosonic representation of the quarks anticommutes.

This case can be treated explicitly for any $N_c$.
Bosonization yields
\bea
&& S = S_{Gauss} + \int d\tau dx (V_{Cartan} + V_{off-diag}), \label{Baluni}\\
&& V_{Cartan} =  -\pi g^2\int dy |x-y|\p_x\Phi_a(\delta_{ab} - 1/N_c)\p_y\Phi_b = \nonumber\\
&&  g^2\Phi_a(\delta_{ab} -1/N)\Phi_b, \label{Kartan}
\eea
where $S_{Gauss}$ is the Gaussian action (\ref{Bose}), we have introduced non-chiral bosons $\Phi = \varphi +\bar\varphi$,  and the
off-diagonal part is given by
\begin{widetext}
\bea
&& V_{off-diag} = \sum_{a>b}\int dy \frac{g^2}{4\pi}|x-y|^{-1} \Big\{\cos[\sqrt{4\pi}[\varphi_{ab}(x) -\varphi_{ab}(y)]] + \cos[\sqrt{4\pi}[\bar\varphi_{ab}(x) -\bar\varphi_{ab}(y)]] -\nonumber\\
&& 2 |x-y|^2\cos[\sqrt{4\pi}[\varphi_{ab}(x) +\bar\varphi_{ab}(y)]]\Big\} \label{offd}
\eea
\end{widetext}
with $\varphi_{ab}=\varphi_{a}-\varphi_{b}$.

The form of the effective potential (\ref{Kartan}) is suggestive of the confinement: since it is not periodic in the fields, it does not allow topological excitations for all field configurations except one.
The only soft mode remaining is the
\bea
\Phi_U = \frac{1}{\sqrt{N_c}}\sum_a \Phi_a,
\eea
which does not enter into $V$. The ground state of (\ref{Baluni}) corresponds to all fields being equal; projecting the mass term on this vacuum we get the sine-Gordon model with renormalized mass term:
\bea
&& {\cal H}_{eff} = \label{sGB}\\
&& \frac{1}{2}\Big\{\Pi^2 +(\p_x\Phi_U)^2\Big\} + 2\frac{\tilde m}{2\pi} \left[1- \cos\left(\sqrt{\frac{4\pi}{N_c}}\Phi_U\right)\right]. \nonumber
\eea
Naturally, the projection assumes that the energy scale generated
by the mass term is much smaller than the energies of the mesonic fields. Even for a single flavor, the complete bosonized
form of QCD involves a set of $N_c$ coupled sine-Gordon models, as first derived by
Baluni \cite{Baluni:1980bw,Steinhardt:1980ry,Cohen:1982mq,Gepner:1984au,Frishman:1992mr,Armoni:1998ny}.
Besides the $U(1)$ field $\Phi_U$, there are also color singlet excitations
above $\Lambda_{QCD}$, involving fluctuations of individual fields $\Phi_a$ around the minimum of the potential.
By going to energies below the scale of the gauge coupling, all of these massive
degrees of freedom can be ignored.  This simplifies the analysis considerably.

\subsection{Low energy theory for $N_f = N_c =2$}  \label{lowenergy22}

 For $N_c = N_f =2$ the indices $\s$ and $f$ take the values, $\s = \pm 1, ~~f = \pm 1$.
 The Gaussian models (Eq. ref{Bose}) in this case have total central charge $c=4$. Our goal is to rearrange the fields into groups with central charges 1, 3/2 and 3/2 corresponding to the U(1), $SU(2)_2$ flavor and $SU(2)_2$ color sectors. The next step is to introduce new bosonic fields, $\varphi_U$, $\varphi_F$, $\varphi_c$, and $\varphi_{cF}$, defined by:
\bea
\varphi_{f,\s} &=&  \frac{1}{2}\Big(\varphi_U +f\varphi_F +\s\varphi_{c} +f\s\varphi_{cF}\Big), \cr\cr
\bar\varphi_{f,\s} &=& \frac{1}{2}\Big(\bar\varphi_U +f\bar\varphi_F +\s\bar\varphi_{c} +f\s\bar\varphi_{cF}\Big).
\eea
This transformation leaves the bosonic action (Eq. \ref{Bose}) covariant. The derivatives of fields $\varphi_U, \bar\varphi_U$ are U(1) currents describing smooth fluctuations of the particle density. From other bosonic fields we construct the other currents - generators of the $SU(2)_2$ flavor and color Kac-Moody algebras. For instance, the color right- and left moving color currents are
\bea
&& J^a_R = \frac{1}{2}\sum_{f=1}^2 R^+_{f,\s}\tau^a_{\s,\s'}R_{f,\s'} = \epsilon_{abc}\chi^b_R\chi^c_R, \nonumber\\
&&  J^a_L = \frac{1}{2}\sum_{f=1}^2 L^+_{f,\s}\tau^a_{\s,\s'}L_{f,\s'} = \epsilon_{abc}\chi^b_L\chi^c_L,
\eea
where $\tau^a$ are Pauli matrices and the labels are $a= (1c,2c,3c)$, are expressed in terms of Majorana fermions. The Majorana fermions describing the color sector  are related to the following chiral bosonic fields via the relations::
\bea
&&\chi_{R,1c} = \frac{1}{\sqrt\pi} \xi_c \sin(\sqrt{4\pi}\varphi_{c})\; , \cr\cr
&& \chi_{R,2c} = \frac{1}{\sqrt\pi} \xi_c \cos(\sqrt{4\pi}\varphi_{c})  \; , \cr\cr
&&\chi_{R,3c}= \frac{1}{\sqrt\pi} \xi_{cF} \cos(\sqrt{4\pi}\varphi_{cF}) \; . \label{Mc}
\eea
The left-moving ones are related similarly to the corresponding $\bar\varphi$-fields.
The Cartan current $J^3$ is understood in the regularized sense, i.e. it corresponds
to derivative of the field $\varphi_c$.
Likewise one can introduce the triad of the Majoranas to represent the $SU(2)_2$ flavor sector:
\bea
&&\chi_{R,1F} = \frac{1}{\sqrt\pi} \xi_F \cos(\sqrt{4\pi}\varphi_{F}) \; , \cr\cr
&& \chi_{R,2F} = \frac{1}{\sqrt\pi} \xi_F\sin(\sqrt{4\pi}\varphi_{F}) \; , \cr\cr
&&\chi_{R,3F}= \frac{1}{\sqrt\pi} \xi_{cF} \sin(\sqrt{4\pi}\varphi_{cF})  .\label{MF}
\eea
Please note that the bosonic field $cF$ takes part in both the color and the flavor sectors since it enters in the Majorana fermions $3c$ and $3F$.
The mass term in terms of the transformed bosons becomes
\begin{widetext}
\begin{eqnarray}
&& \sum_{f,\s =\pm 1}(R^+_{f,\s}L_{f,\s} + H.c.) =
2\sum_{f,\s =\pm 1}\cos\Big[\sqrt\pi(\Phi_U + f\Phi_F + \s\Phi_c +f\s\Phi_{cF})\Big]
\nonumber \\
&&=8\Re e\Big\{\re^{i\sqrt\pi \Phi_U}\Big[\cos(\sqrt\pi\Phi_F) \cos(\sqrt\pi\Phi_{cF})\cos(\sqrt\pi\Phi_{c}) - i \sin(\sqrt\pi\Phi_F) \sin(\sqrt\pi\Phi_{cF})\sin(\sqrt\pi\Phi_{c}) \Big]\Big\} \nonumber\\
&& \sim \Re e\Big\{\re^{i\sqrt\pi \Phi_U}\Big[(\mu_{1F}\mu_{2F}\mu_{3F})(\mu_{3c}\mu_{1c}\mu_{2c})
-i (\s_{1F}\s_{2F}\s_{3F})(\s_{3c}\s_{1c}\s_{2c})\Big]\Big\} \; .
\label{bilinear}
\end{eqnarray}
\end{widetext}
In the last line we used the fact
that the cosine and sine fields of $\sqrt\pi\Phi$ can be
expressed in terms of the Ising model order and disorder parameter fields $\s$ and $\mu$,  as explained in Appendix
\ref{app_ising}.

Now we can proceed further since the Ising fields can be used to represent the $SU(2)_2$ spin-1/2 primary ones.
For example, the spin-1/2 matrix field in the flavor sector can be represented as \cite{Zamolodchikov:1986bd}
\bea
&&\hat G_{\alpha\beta} = \hat I \s_{1F}\s_{2F}\s_{3F} \label{Gmatrix} \\
&&+i\Big(\hat\tau^1 \mu_{1F}\s_{2F}\s_{3F} + \hat\tau^2 \s_{1F}\mu_{2F}\s_{3F} +\hat\tau^3\s_{1F}\s_{2F}\mu_{3F}\Big)
\; .
\nonumber
\eea
where $\hat I$ is the identity and $\hat\tau^a$ are the Pauli matrices. Since at the critical point the Ising model is self dual, one can interchange  $\s_a$'s and  $\mu_a$'s in Eq. (\ref{Gmatrix}) which results in the alternative (dual) representation $\hat{\tilde G}$. Although correlation functions of $\hat G$ and $\hat{\tilde G}$ are identical, these operators are nonlocal with respect to each other  and as such cannot be present in the low energy action simultaneously.

Looking at Eq. (\ref{bilinear}), we see that the fermionic bilinears do not factorize into products of single valued SU(2)$_2$ WZNW fields because they contain mutually nonlocal fields. As we mentioned above, this was noticed already in Ref. \cite{Frishman:1992mr} for the case of general  $N_c,N_f$. The factorization occurs for conformal blocks, not for single valued operators.  Although the present discussion includes only a particular case, it is intended as an illustration of the general statement, namely, that the projection to the strong coupling vacuum chooses one particular representation.  This follows from the fact that the interaction of the color currents spontaneously breaks the symmetry between the $\s$ and $\mu$ vacua such that either $\s$'s or $\mu$'s condense (they cannot condense together). It is straightforward to check this for the NJL model where the point-like current-current interaction can be decoupled by the Hubbard-Stratonovich transformation:
\bea
\frac{g}{2}(\chi_R^a\chi_L^a)^2 \rightarrow \Delta^2/2g + i\Delta(\chi_R^a\chi_L^a).
\eea
Then one can show that integration over the fermions produces a double-well  effective potential for $\Delta$. In the ground state the $Z_2$-symmetry between the minima is spontaneously broken. Now recall that the Ising model is equivalent to massive Majorana fermion and the sign of the mass determines which of the fields $\s$ or $\mu$  condenses.
We suggest that the same mechanism works for QCD, which differs from the NJL model only
in that the interaction in QCD is long range.
The long range character of the interaction should only strengthen symmetry breaking.

We elaborate further on the projection argument. In the limit of vanishing mass,
the eigenstates of our theory are tensor products of the states of the color and the flavor and $U(1)$ sectors.
The color and flavor states  are created respectively by Majorana fermions, Eqs. (\ref{Mc}) and (\ref{MF}),
and their left-moving counterparts. As we have said, the vacuum in the color sector is double degenerate.
In one vacuum, $|0_{c\s}\ra$, we have
 \bea
\la \s_{1c}\s_{2c}\s_{3c}\ra \neq 0 \; ,
\eea
while in the other, $|0_{c\mu}\ra$,
 \bea
\la \mu_{1c}\mu_{2c}\mu_{3c}\ra \neq 0 \; ,
\eea
 At $T=0$ the symmetry is spontaneously broken and we have to project the mass term to one of the vacua, say $|0_{c\s}\ra$ .

 Suppose that vacuum of the theory is $|0_c\rangle$.  Let then $|f,0_{c\s}\rangle = f^\dagger_1\cdots f^\dagger_n|0\ra\otimes|0_{c\s}\rangle$ with $f^+$ being positive momentum components of the flavor Majoranas (\ref{MF}), be a shorthand notation for a flavor excitation above the vacuum.  If $\la 0_{c\s}| \s_{1c}\s_{2c}\s_{3c}|0_{c\s}\ra $ is non-zero then
\bea
&& \la f,0_{c\s}| (\mu_{1F}\mu_{2F}\mu_{3F})(\mu_{3c}\mu_{1c}\mu_{2c}) |f',0_{c\s}\ra=0 \cr\cr
&& \la f,0_{c\s}|(\s_{1F}\s_{2F}\s_{3F})(\s_{3c}\s_{1c}\s_{2c})|f',0_{c\s}\ra \cr\cr
&& = \la 0_{c\s}| \s_{1c}\s_{2c}\s_{3c}|0_{c\s}\ra \la f|(\s_{1F}\s_{2F}\s_{3F})|f'\ra
\eea
for states with arbitrary flavor excitations (i.e. $f\neq f' \neq 0$) above the color singlet vacuum.   The projection to the color singlet cuts from the mass term the single valued field $\hat G$  acting in the flavor sector with matrix elements between different $f$ and $f'$.
Hence the mass term projected on this ground state is
\begin{equation}
m(R^+_{f,\s}L_{f,\s} + H.c.) \longrightarrow \frac{\tilde m}{4\pi} \Big(\re^{i\sqrt\pi \Phi_U}
\mbox{Tr}\hat G + H.c.\Big) \; ,
\end{equation}
where according to (\ref{Gmatrix}) $\mbox{Tr} \hat G = 2\s_{1F}\s_{2F}\s_{3F}$
and $
\tilde m \propto  m\la \s_{1c}\s_{2c}\s_{3c}\ra.$

The above example illustrates our point that local fields emerge as a result of projection
on the singlet ground state of gapped quarks. 
From this  derivation we conjecture that  for general
$N_c$ and $N_f$  the strong coupling  Lagrangian is \cite{Gepner:1984au,Frishman:1992mr}
\bea
{\cal L} &=& \frac{1}{2}(\p_{\mu}\Phi_U)^2 + W[SU(N_f)_{N_c};G] \cr\cr
&+& \frac{\tilde m}{4\pi} :\Big(\re^{i\sqrt{4\pi/N_cN_f}\Phi_U}\mbox{Tr}G + H.c.\Big): \; ;  \label{NABQCD11}
\eea
where $G$ is the spin-1/2 primary field of the SU$_{N_c}(N_f)$ WZNW model. $::$ denotes normal ordering introduced in such a way that the theory has a new ultraviolet cut-off
$\Lambda_{QCD} \sim g$.
This means that the two-point correlation function of the operator ${\cal O}_{h,\bar h}$ with conformal dimensions $(h,\bar h)$ is equal to
\bea
&& \la\la {\cal O}(\tau,x){\cal O}(0,0)\ra\ra = \label{twoptfnnormCFT}\\
&& [(\tau -i x)]^{-2h}[(\tau + i x)]^{-2\bar h},\nonumber
\eea
with $\tilde m$ being the renormalized quark mass at the scale of $\Lambda_{QCD}\sim g$:
\bea
\tilde m \sim m_{ren} g^{d+\frac{1}{N_c N_f}}, ~~d = \frac{N_c-1/N_c}{N_c+N_f},
\eea
with $d$ being the scaling dimension of the $SU(N_c)_{N_f}$ WZNW matrix field.

\section{Matter at nonzero density}   \label{nonzerodensity}

At nonzero chemical potential $\mu$ one may shift the U(1) field
\bea
\Big(4\pi/N_cN_f\Big)^{1/2}\Phi_U \rightarrow 2k_0 x + \Big(4\pi/N_cN_f\Big)^{1/2}\Phi_U. \label{finitedensity}
\eea
so that $\mu=N_c N_f k_0$.
It is expected that the presence of the oscillations makes the mass term irrelevant.
 In this case we have a ``strange'' metal: both the charge and flavor excitations are gapless,
and all correlation functions of baryons have a power law decay, with non-trivial scaling exponents.
This is true for all values of $N_f$ and $N_c$ for the standard QCD Lagrangian.
In Ref. \cite{Azaria:2016mqb}, a generalized NJL model was considered
possessing a current-current interaction in the U(1) sector.  In the
presence of such an interaction, a fusion of operators at second order
in $\tilde m$ may generate quantities where the oscillatory terms
cancel and produce a relevant perturbation proportional to ${\rm Tr}\Phi_{adj}$, where $\Phi_{adj}$ is a WZNW primary field in the adjoint representation
of the flavor symmetry group.  This perturbation, in this more generalized setting, gaps out the flavour sector, and the model's ground state is not a metal.
We here however will not consider the addition of this U(1) current-current perturbation to the theory.

In this section we will focus on describing the strange metal that results at non-zero density.  This metal has a Luttinger liquid description described by the following effective low-energy bosonic action:
\bea
{\cal L}_{eff} = \frac{\tilde{K}(\mu)}{2}\Big[v_F(\mu)^{-1}(\p_{\tau}\Phi)^2 + v_F(\mu)(\p_x\Phi)^2\Big].
\label{Luttinger}
\eea
This action depends only upon 
the Luttinger parameter, $\tilde{K}(\mu)$, and the Fermi velocity, $v_F(\mu)$.
In the following two subsections, we determine these parameters as a function of $\mu$
in two cases.  First, for a single flavor, $N_f = 1$, and an arbitrary number of colors, $N_c \geq 2$, where we have
recourse to exact methods.  Second, in the case of arbitrary flavors and colors, $N_f, N_c \geq 1$,
by using perturbative methods appropriate for the case when $\mu$ far exceeds the mass gap in the $\mu=0$ theory.
These two parameters determine the low energy behavior of correlation functions.
In the final subsection here we will give an example of this in the specific case of $N_c=3$ and $N_f=2$.

\subsection{The case of $N_f=1$}     \label{Nf1TBA}

We begin by determining $\tilde{K}(\mu)$ and $v_F(\mu)$ for $N_f=1$ as this case allows for an exact treatment in strong coupling.
As was discussed in Section I.A, the resulting low energy Hamiltonian is a sine-Gordon model (\ref{sGB}):
\be
{\cal L} = \frac{1}{2}(\p_{\mu}\Phi_U)^2 - \frac{\tilde m}{2\pi}\cos(\sqrt{\frac{4\pi}{N_c}}\Phi_U+2k_0 x).\label{beta}
\ee
This model is exactly solvable and provides insight into more realistic cases.  The spectrum of this model consists of a soliton($s$)/anti-soliton($\bar s$) pair and $n=1,\cdots,N_c$ breathers.   All of these excitations are color singlets.
The soliton and antisoliton carry $\pm 1$ U(1) charge and have mass $m_s$, while the breathers are U(1) neutral and have mass 
\begin{equation}
m_n=2m_s\sin(\pi n \xi /2), ~~ n=1,\cdots,N_c=\nu-2,
\end{equation}
with $\xi=1/(2N_c-1)$.
We first consider what happens to these masses when $\mu > m_s$, answering the fundamental question of what the ground state is of this model at finite density.  It turns out that it is metallic, as advertised, consisting of a Fermi sea purely of solitons.

\begin{figure}[!htb]
\includegraphics[width=\linewidth,clip]{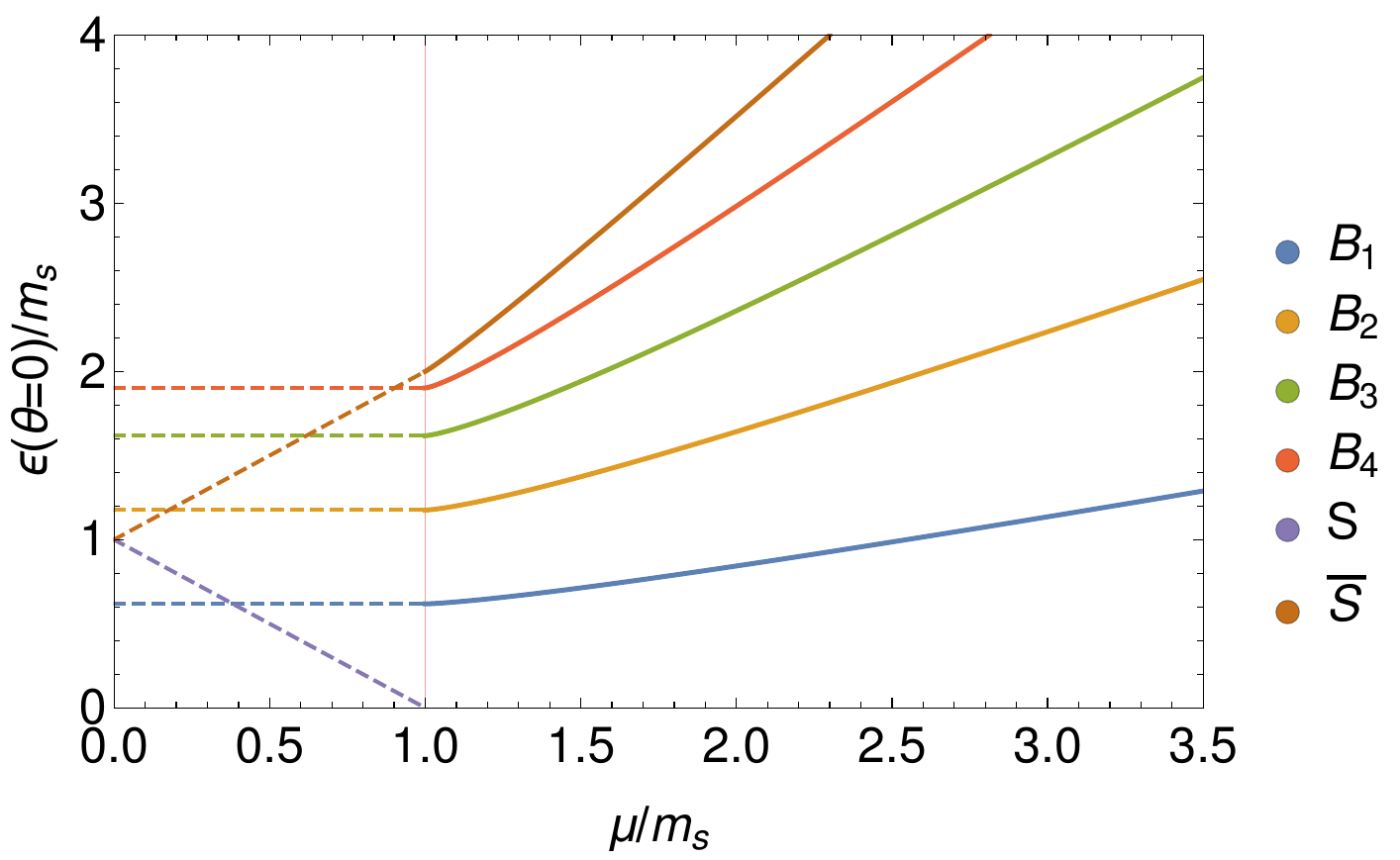}
\caption{Spectral gaps of the soliton ($S$), anti-solition ($\overline{S}$), and the four
   breathers, $B_1 \ldots B_4$, versus the chemical potential. Here $N_c=3$, $N_f=1$,
   calculated from Eqs. \eqref{tba1}--\eqref{tba1v}.
  The spectral gap of the solition is the mass minus the chemical potential, and vanishes
when $\mu = m_s$.
}
\label{fig_mass}
\end{figure}

To describe what happens to the spectrum when $\mu>m_s$, we employ the following set of thermodynamic Bethe ansatz (TBA) equations.   These equations describe the energies, $\epsilon_{s,\bar s},\epsilon_n$ of the different excitations.  These equations, as written below, presume that only the solitons appear in the ground state.  If at a given rapidity $\theta$, $\epsilon(\theta)<0$, the particle appears in the ground state at zero temperature.  If $\epsilon(\theta) > 0$ then the particle is excluded from the ground state.  The basic idea is that we show these equations are self-consistent, i.e., that if only solitions are supposed to appear in the ground state then we find that $\epsilon_{\bar s},\epsilon_n$ are alway positive.  The TBA equations are given by \cite{Japaridze:1984dz}:
\begin{widetext}
\bea
\epsilon_s(\theta,\mu) &+& \int_{-B}^B \mathcal{K}_{ss}(\theta-\theta^\prime)\epsilon_s(\theta^\prime,\mu) \rd \theta^\prime = m_s\cosh \theta - \mu, \quad \epsilon_s(B,\mu) =0; \label{tba1} \\
\epsilon_{\bar{s}}(\theta,\mu) &+& \int_{-B}^B \mathcal{K}_{\bar s s}(\theta-\theta^\prime)\epsilon_s(\theta^\prime,\mu) \rd \theta^\prime = m_s\cosh \theta + \mu \; ; \\
\epsilon_n(\theta,\mu)&+&\int_{-B}^B \rd \theta^\prime \mathcal{K}_{ns}(\theta-\theta^\prime) \epsilon_s(\theta^\prime,\mu) = m_n\cosh \theta \; ; \\
~~\mathcal{K}_{ss}(\omega) &=& \frac{\sinh[(1 -1/(\nu -1))\pi\omega/2]}{2\sinh[\pi\omega/(2(\nu -1))]\cosh(\pi\omega/2)}\; ;\quad \mathcal{K}_{ns}(\omega) = \frac{\coth[\pi\omega/(2(\nu -1))]\sinh[\pi n\omega/(2(\nu -1))]}{\cosh(\pi\omega/2)} \; \label{tba1v}.
\eea
\end{widetext}
Here $\nu \equiv N_c+2$ and the rapidity $\theta$ parameterizes energy-momentum (E/p) of a particle of mass, $m$, via
\begin{equation}
E=m\cosh(\theta); ~~~ p = m\sinh(\theta).
\end{equation}
The parameter $B$ determines the Fermi momentum through $k_F=m_s\sinh(B)$,
 -- see Fig. (\ref{FigBdep}) in Appendix \ref{app_num}. Note that $k_F$ is the dressed version of the 'bare' Fermi momentum, $k_0$, introduced in \eqref{finitedensity}.
More details on the TBA equations are presented in Appendix \ref{app_tba}.
In the above,
$\mathcal{K}_{ij}(\omega)$ are the Fourier transforms of $K_{ij}(\theta)$
with respect to $\theta$.  The soliton mass can be related to $\tilde m$ -- see Ref. \cite{Zamolodchikov:1995xk}.
When $\mu$ exceeds the soliton mass, the soliton spectrum becomes gapless, while
the spectral gaps of the anti-soliton and the breathers start to increase, as illustrated
in Fig. (\ref{fig_mass}) for $N_c=1$.  This demonstrates that it is only the soliton that appears in the ground state and that interactions between the breathers and the Fermi sea of solitons do not lead to the breathers themselves becoming gapless.

Now that we have characterized the finite density ground state, let us turn to determining the Luttinger parameter, $\tilde K$, and the Fermi velocity $v_F$.
The Luttinger parameter can be extracted \cite{Izergin:1988kw, PhysRevB.63.085109}
by solving the equation for the so-called dressed charge, $\zeta(\theta)$:
\bea
\zeta(\theta) + \int_{-B}^B {\cal K}_{ss}(\theta-\theta^\prime)\zeta(\theta^\prime) \rd \theta^\prime = 1.  \label{zetaTBA}
\eea
The dressed charge measures the change in the charge of the system when one soliton is added to the system while keeping $B$ fixed.  The dressed charge is not unity as interactions lead to some of the rapidities of solitons in the Fermi sea exceeding $B$ and so being expelled from the ground state (in the sense of the grand canonical ensemble).
The Luttinger parameter is given in terms of the dressed charge as follows:
\bea
 \tilde{K}(\mu) =\zeta^2(B).\label{Kc}
\eea
The numerical analysis of this equation is presented in Appendix \ref{app_num}.
In the limit of large chemical potential we can however write down the leading order solution:
at $\theta=B$:
\begin{equation}
\zeta(B) = [1+{\cal K}_{ss}(\omega =0)]^{-1/2}=(\nu/2)^{-1/2}=N_c^{-1/2}.
\end{equation}
As the Fermi surface shrinks, so that $B \rightarrow 0$, $\zeta$ approaches unity as $\zeta=1-2^{3/2}{\cal K}_{ss}(\theta =0)\delta_\mu^{1/2}+O(\delta_\mu)$ with $\delta_\mu=\mu/m_s-1$. 
In fact for any number of flavors, $\tilde{K}(\mu) \rightarrow 1$ as
the Fermi surface vanishes, $B \rightarrow 0$.

\begin{figure}[!htb]
\includegraphics[width=\linewidth,clip]{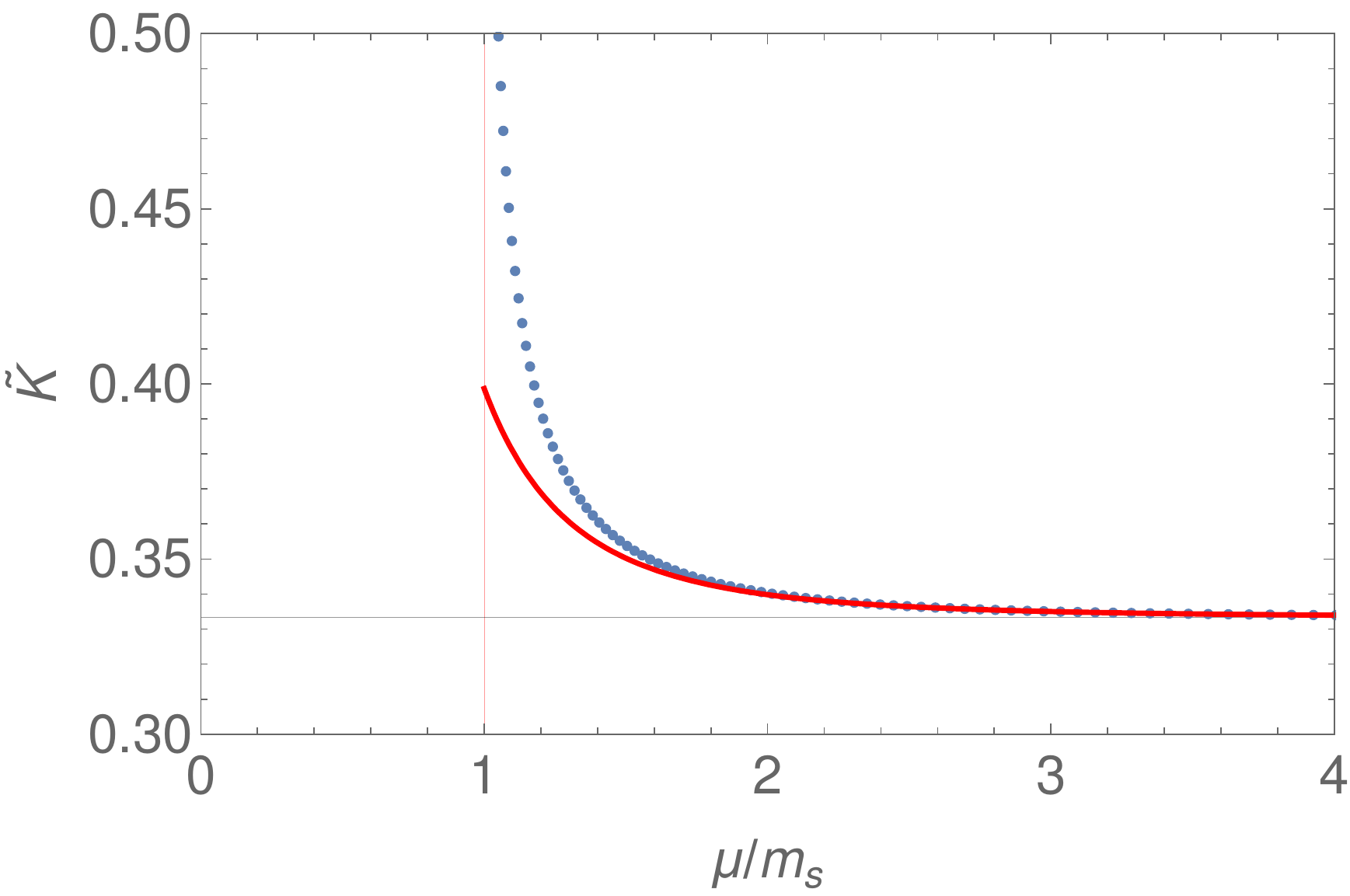}
\caption{The Luttinger parameter $\tilde{K}$ as a function of $\mu$, divided by the soliton mass, for the model with
  $N_f=1$ and $N_c=3$. The solid (red) curve is the result to leading order in perturbation theory (Eq. \eqref{leadingK}),
  the dotted (blue) curve from the solution of the
  thermodynamic Bethe ansatz in Appendix (\ref{app_num}).
  The latter shows that this is a  Luttinger liquid for {\it any} $\mu > m_s$.
\label{Spec}}
\end{figure}

\begin{figure}[!htb]
\includegraphics[width=\linewidth,clip]{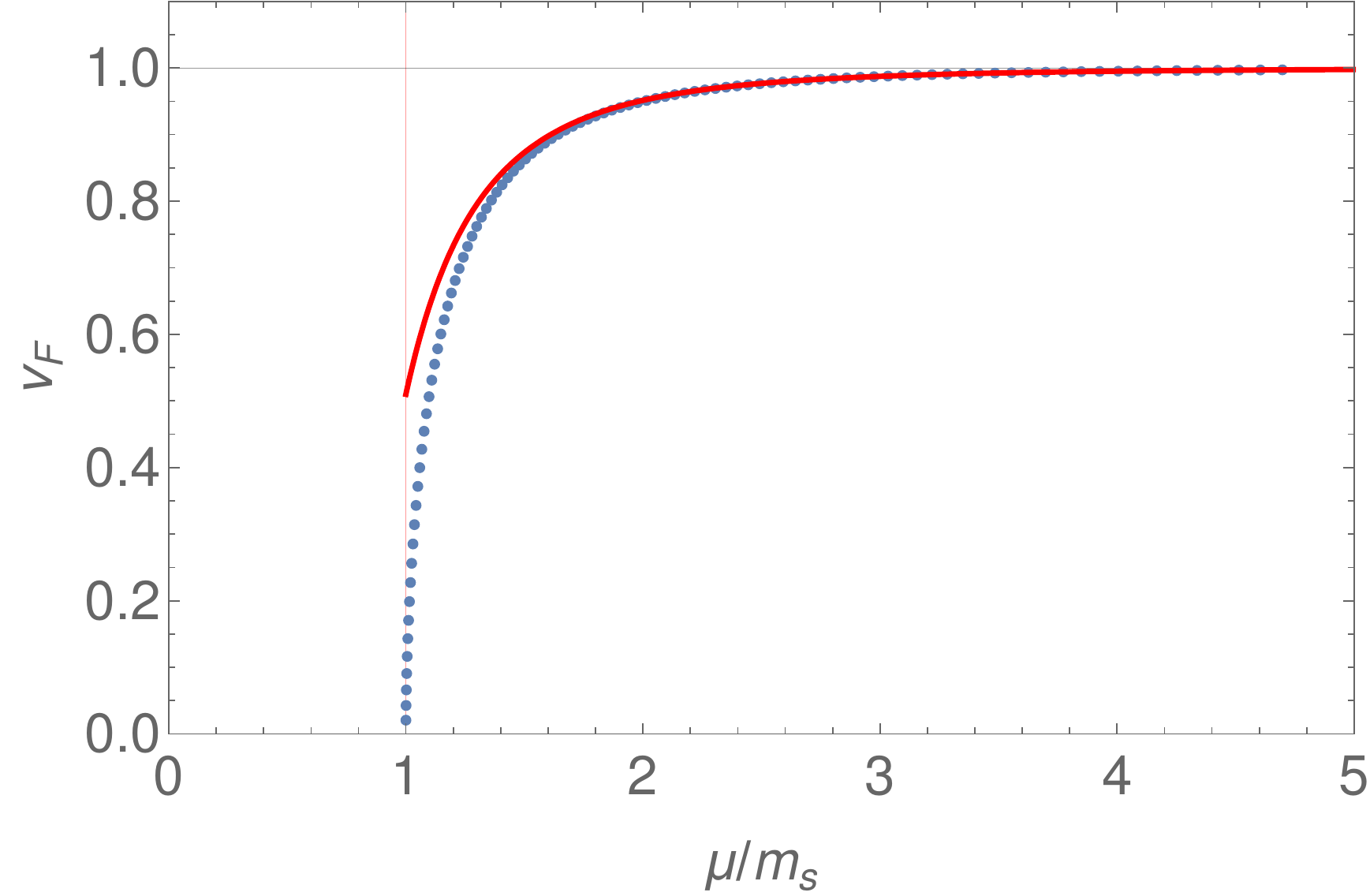}
\caption{ The Fermi velocity, Eq. (\ref{FermiVF}), versus the chemical potential.
  The solid (red) curve is the result to leading order in perturbation theory in the mass (Eq. \eqref{leadingvF});
  the dotted (blue) curve from the solution of the thermodynamic Bethe ansatz in
  Appendices (\ref{app_tba}) and (\ref{app_num}).   }
 \label{fig_vF}
\end{figure}

With the relation $B(\mu)$ in hand, it is then possible to compute the Fermi velocity $v_F$ via
\begin{equation}
    v_F=\left.\frac{\partial \epsilon_s(\theta,\mu)}{\partial \theta}\right|_{\theta=B}\frac{1}{2\pi\rho_s(B)},
    \label{FermiVF}
\end{equation}
where the rapidity density distribution function $\rho_s(\theta)$ obeys the same equation as Eq. (\ref{tba1}), with
the term $m_s\cosh\theta- \mu$ replaced by $(2\pi)^{-1}m_s\cosh\theta$ on the right hand side.  The behavior of the Fermi velocity versus $\mu$ is plotted in
Fig. (\ref{fig_vF}): it vanishes at threshold, when $\mu = \epsilon_s$, and approaches unity at
asymptotically high density.

To summarize, for a single flavor, by projecting onto the color singlet sector, we obtain an effective theory
involving only $U(1)$ fields, the sine-Gordon model of Eq. (\ref{beta}).  At zero chemical potential,
all particles are massive: both the baryons (i.e., solitons and anti-solitons)  and the mesons (the breathers).  
At finite density, however, there is an interacting Fermi sea composed solely of baryonic (solitonic)
particles.  As we show in Section \ref{Strangemetal32}, this interacting Fermi sea leads to correlation functions characterized by power laws which are a function of the Luttinger parameter, $\tilde K(\mu)$.

\subsection{Perturbation theory in the mass}     \label{PTsubsec}

The problem for general $N_f$ and $N_c$ is not integrable.
The presence of the mass term changes both
the velocity of the $SU(N_f)$ and $U(1)$ excitations as well as the Luttinger parameter $\tilde{K}$ in the $U(1)$ sector.
Here we focus on the $U(1)$ sector.

At high density, one can compute both $\tilde{K}$ and the velocity of the $U(1)$ excitations, $v_F$,  by expanding in powers of the mass.
There are several ways to determine these parameters using perturbation theory.
We here focus on computing the charge susceptibility, which is the Fourier transform,
\begin{equation}
    \chi(q,\Omega)=\intop dx \; dt \; {\rm e}^{i q x+i \Omega t} \chi(x,t) \; , \label{susc0}
\end{equation}
of the two-point function of
$J_0$, the time component of the $U(1)$ current, 
\begin{equation}
    \chi(x_1-x_2,t_1-t_2)=i \Theta(t_1-t_2)\left\langle\left[J_0(x_1,t_1),J_0(x_2,t_2)\right]\right\rangle
\end{equation}
as given by the usual Kubo formula.
It is easy to calculate the susceptibility for imaginary frequencies using the Lehmann spectral expansion,
which can be written as a $\tau$-ordered product
\begin{eqnarray}
        \chi(q,\Omega=i\omega) &=&\intop_{-\infty}^\infty dx d\tau e^{iqx+i\omega \tau}\cr\cr
&& \hskip 0.2in \times \mathcal{T}_\tau\left\{\left\langle0\left|J_0(x,\tau)J_0(0,0)\right|0\right\rangle\right\}, \label{susceucl}
\end{eqnarray}
where $J_0(x,\tau)\equiv J_0(x,t=-i\tau)$.

Let us first recall the exact expression for the U(1) susceptibility in the effective theory Eq. \eqref{Luttinger}.
The bosonic field of the effective theory admits the mode expansion
\bea
   &&\Phi(x,t)=\Phi_0+ L^{-1}\Pi_0 t+\label{cylphi}
\\
   && \frac{\sqrt{v_F}}{\sqrt{2L\tilde{K}}}\sum_{n\neq0}\frac{1}{\sqrt{\omega_n}}\left(a_n e^{i k_n x-i\omega_n t}+a^\dagger_n e^{-i k_n x+i\omega_n t}\right),\nonumber
\eea
where
\begin{equation}
    k_n=\frac{2 \pi n}{L},\quad \omega_n=v_{F}\left|k_n\right|.
\end{equation}
In the effective theory, the U(1) current is given as
\bea
    J_\mu(x,t)=-\frac{\tilde{K}}{\sqrt{\pi}}\epsilon_{\mu\nu}\partial_\nu \Phi(x,t) \label{currentdef}
\eea
which ensures that the corresponding charge operator has unit weight on asymptotic states \cite{KonikFendley}.
Differentiating with respect to $x_i$, performing the sum, and taking the limit $L\rightarrow\infty$, we obtain for the correlator
\bea
  && \mathcal{T}_\tau\left\{\left\langle0\left|J_0(x,\tau)J_0(0,0)\right|0\right\rangle_0\right\} =\nonumber\\
  && -\frac{\tilde{K}}{4\pi^2}\left(\frac{1}{(x+iv_{F0}\tau)^2}+\frac{1}{(x-iv_{F0}\tau)^2}\right).
\eea
Straightforward integration then provides the Fourier transform as,
\bea
  \chi(q,i\omega)=\frac{v_F \tilde{K}}{\pi}\frac{q^2}{\omega^2+v_F^2 q^2}.     \label{chi1}
\eea
We expect that the above form of the the susceptibility holds for strong-coupling, low energy QCD as long as $q \ll 2k_F$.
The static susceptibility is given by the subsequent limits $\lim_{q\rightarrow0}\left(\lim_{\omega\rightarrow0} \chi(q,\omega)\right)$:
\begin{equation}
    \chi(\omega=q=0) =\frac{\tilde{K}(\mu)}{ \pi  v_F(\mu)} .\label{suscchi}
\end{equation}
We now turn to computing the same quantity in perturbation theory in $\tilde m$ based on the undoped Hamiltonian \eqref{NABQCD11}.  By comparing the result, we will thus be able to derive an expression for $\tilde K(\mu)$ and $v_F(\mu)$ in terms of $\tilde m$.

To proceed, we add a chemical potential to \eqref{NABQCD11} and compute the susceptibility to leading order in $\tilde m^2$.
For small $q,\omega$, the expansion of the susceptibility is necessarily also an expansion around $\mu=\infty$ and can be written as
\be
\chi=\chi_0+\chi_1+\dots.  \label{chiexpand}
\ee
In the language of the undoped Hamiltonian, the $U(1)$ current, written as $J^{undoped}_\mu$ in order to distinguish it from the current operator of the effective theory,
is given by
\bea
    J^{undoped}_\mu(x,t)=-\frac{1}{\sqrt{\pi N_c N_f}}\epsilon_{\mu\nu}\partial_\nu \Phi_U(x,t) \label{currentdefB}
\eea
where the mode expansion of the field $\Phi_U$ has the same structure as $\Phi$ in Eq. \eqref{cylphi}, with the substitutions
$v_F\rightarrow 1$, $\tilde{K}\rightarrow1$

With the current $J^{undoped}$, the zeroth order term of the susceptibility now takes the form
\bea
  \chi(q,i\omega)_0=\frac{1}{N_cN_f\pi}\frac{q^2}{\omega^2+q^2}     \label{chi0}
\eea
This form is valid for the system as $\mu \rightarrow \infty$.  We thus immediately see
\begin{eqnarray}\label{zeroth}
\tilde K(\mu=\infty) &=& (N_cN_f)^{-1};\cr\cr
v_F(\mu=\infty) &=& 1.
\end{eqnarray}
This generalizes our finding from Section \ref{Nf1TBA} that $\tilde K(\mu=\infty) = 1/N_c$ where $N_f=1$.

Let us now proceed to the first nonvanishing perturbative correction. The Euclidean two-point function is formally written as
\begin{eqnarray}
    \left\langle J_0(x_a,\tau_a) J_0(x_b,\tau_b)\right\rangle&=&\cr\cr
&&\hskip -1in \frac{\left\langle \mathcal{T}_\tau J_0(x_a,\tau_a) J_0(x_b,\tau_b)S\right\rangle_0}{\left\langle S\right\rangle_0}.
    \label{eq3}
\end{eqnarray}
where
\begin{eqnarray}
    S&=&\sum_{n=0}^\infty \frac{(-1)^n}{n!}\intop_{-\infty}^\infty\dots\intop_{-\infty}^\infty d\tau_1\dots d\tau_n \cr\cr
&& \hskip 1in \times \mathcal{T}_\tau \left\{ V(\tau_1)\dots V(\tau_n)\right\}
    \label{eq4}
\end{eqnarray}
and $V$ is a short-hand for the mass term in Eq. \eqref{NABQCD11} and $\mathcal{T}_\tau$ refers to euclidean time ordering.

To get the $\tau$-ordered correlator, it is sufficient to compute the expression where the arguments are already in order, $\tau_1>\tau_2>\dots>\tau_n$.
Analytical continuation in the $\tau$ variables then automatically provides the $\tau$-ordered quantity.
In the following, we will consider the following normalization of the WZNW two-point function:
\begin{equation}
    \left\langle G_{\alpha_1}^{\beta_1}(z,\bar{z})G^{-1\:\alpha_2}_{\beta_2}(0,0)\right\rangle=N_f^{-1}\delta_{\alpha_1}^{\alpha_2}\delta_{\beta_1}^{\beta_2}(z\bar{z})^{-2\Delta}
\end{equation}
where $\Delta$ is the chiral dimension of the $SU_{N_c}(N_f)$ WZNW chiral field $G$,
\be
\Delta=\frac{N_f^2-1}{2N_f(N_c+N_f)}.    \label{defDelta}
\ee
The normalization is chosen such that the two-point function of $\mathrm{tr}\:{G}$ is consistent with \eqref{twoptfnnormCFT}.
Other normalization choices would only affect the relation between $\tilde{m}$ and the bare parameters,
the quark mass $m$ and the gauge coupling $g$.
The leading correction reads:
\begin{widetext}
\begin{eqnarray}
 &&  \left\langle J_0(x_a,\tau_a)J_0(x_b,\tau_b)\right\rangle_1=\frac{1}{2}\left(\frac{\tilde{m}}{4\pi}\right)^2 \frac{1}{\pi N_c N_f} \label{eq5v}
\intop_{-\infty}^\infty\intop_{-\infty}^\infty \rd\tau_1 d\tau_2
\intop_{-\infty}^\infty\intop_{-\infty}^\infty \rd x_1 \rd x_2(z_1-z_2)^{-2\Delta}(\bar{z}_1-\bar{z}_2)^{-2\Delta}\\  
 &&\bigg\langle\partial_x\phi(x_a,\tau_a)\partial_x\phi(x_b,\tau_b)\left(e^{i\beta\phi(x_1,\tau_1)}\re^{-i\beta\phi(x_2,\tau_2)}e^{2i k_0(x_1-x_2)}+e^{-i\beta\phi(x_1,\tau_1)}\re^{i\beta\phi(x_2,\tau_2)}e^{-2i k_0(x_1-x_2)}\right)\bigg\rangle_c \nonumber
  \end{eqnarray}
\end{widetext}
where $\beta=\sqrt{4\pi/(N_c N_f)}$, $z_{1,2}=\tau_{1,2}-i x_{1,2}$ and the subscript $c$
stands for the connected component of the correlator.

The mass term is a relevant operator for all $N_c$ and $N_f$. For $N_c=1$, the dimension is $1$ for any $N_f$ as the level 1 WZNW
model corresponds to free fermions. When $N_f>1$,
the scaling dimension decreases with increasing $N_c$ and attendantly the perturbation becomes more relevant.
For a fixed $N_c$, the perturbation becomes less relevant by increasing $N_f$.
Since the perturbation is always strongly relevant for $N_c > 1$, no ultraviolet divergences appear at finite $k_0$ at second order.
The derivation of bosonic expectation values of this type is given in Appendix (\ref{app_details}).
Introducing the parameter,
\begin{equation}
    \alpha=\left(\frac{\beta^2}{4\pi}+2\Delta\right)=\frac{1+N_c N_f}{N_c^2+N_c N_f} \; , \label{alphadef}
\end{equation}
our calculation gives
\begin{widetext}
\begin{eqnarray}
\chi(q,i\omega)_1 &=&
2\left(\frac{\tilde{m}}{4\pi}\right)^2 k_0^{-2+2\alpha}  \frac{1}{N_c N_f} \beta^2   \frac{q^2}{(q^2+\omega^2)^2}\frac{\Gamma(1-\alpha)}{\Gamma(\alpha)} \times \nonumber \\
&& \left\{- 1+ \frac{1}{2}\left[\left(\frac{\omega^2}{(2k_0)^2}+\left(-1+\frac{q}{2k_0}\right)^2 \right)^{-1+\alpha}+\left(\frac{\omega^2}{(2k_0)^2}+\left(1+\frac{q}{2k_0}\right)^2 \right)^{-1+\alpha}\right]\right\} \; .\label{PertResult}
\end{eqnarray}
\end{widetext}

\begin{figure}[!htb]
\includegraphics[width=\linewidth,clip]{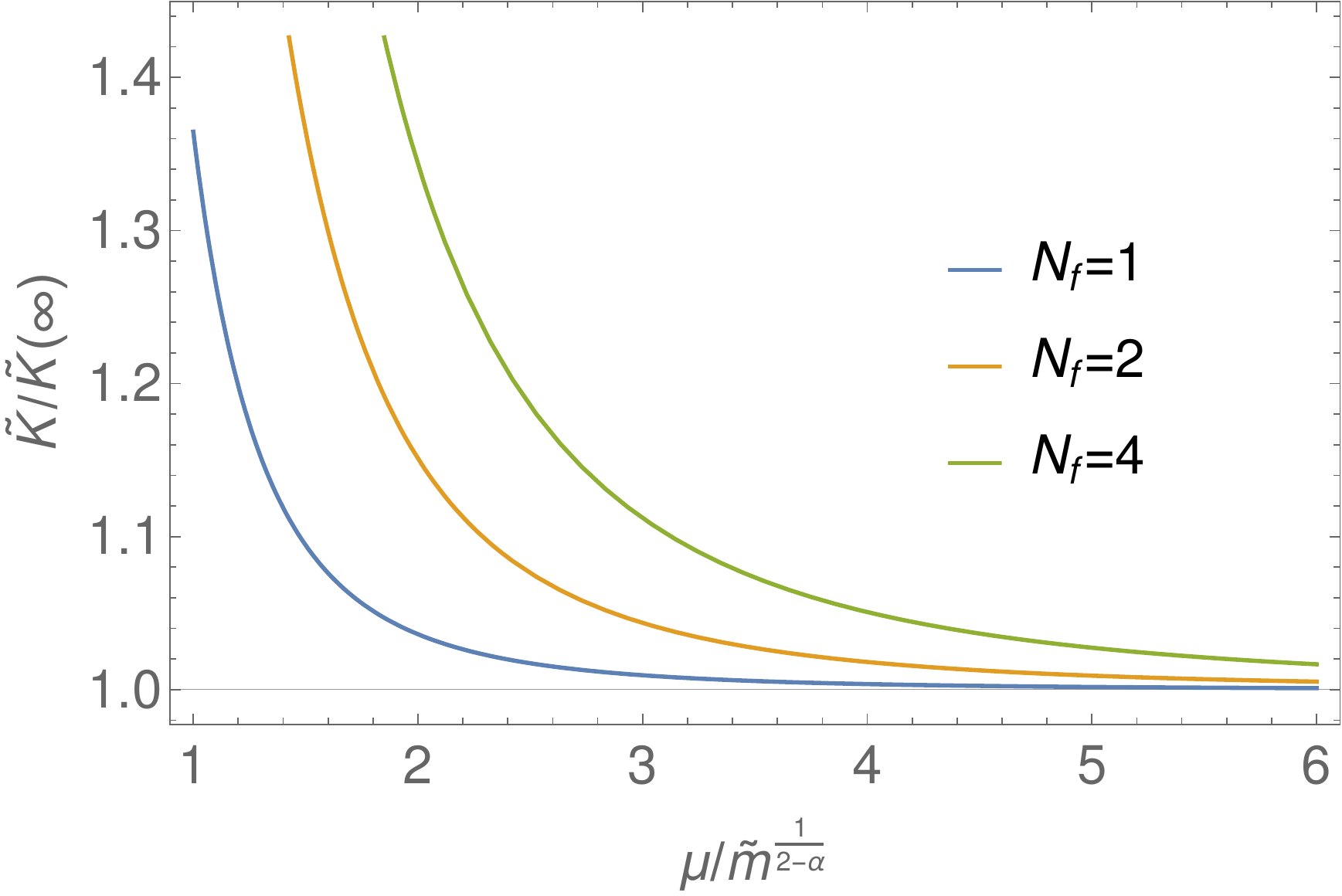}
\caption{Here are plotted the Luttinger parameter $\tilde{K}(\mu)/\tilde{K}(\infty)$, $\tilde{K}(\infty)=(N_c N_f)^{-1}$ as a function of the dimensionless
ratio $\mu/\tilde{m}^{1/(2-\alpha)}$, for the models with
$N_c=3$ and different values of $N_f$ to leading order in perturbation theory \eqref{leadingK}.}
 \label{FigPertSpec} 
\end{figure}

The Luttinger parameter and the velocity can be obtained separately
by a careful examination of the $q\rightarrow0$, $\omega\rightarrow0$ limits.
First we formally expand the parameters $\tilde{K}$ and $v_F$ in Eq. \eqref{chi1} with
respect to a small parameter $\epsilon$ 
\begin{eqnarray}
    \tilde{K}&=&\tilde{K}(\mu=\infty)+\epsilon \tilde{K}_1+\dots\label{leadingK}\\
    v_F&=&v_{F0}+\epsilon v_{F1}+\dots\label{leadingvF}
\end{eqnarray}
defined as
\be
\epsilon=k_0^{-4+2\alpha}\left(\frac{\tilde{m}}{4\pi}\right)^2\frac{2\pi\beta^2}{4}\frac{\Gamma(1-\alpha)}{\Gamma(\alpha)}.
\ee 
It is convenient to focus our attention on the lines $\omega=\delta q$,
$\delta$ being a free parameter. In this case Eq. \eqref{chi1} becomes independent of $q$.
Implementing the same substitution to the expansion Eqs. \eqref{chiexpand}, \eqref{chi0}, and \eqref{PertResult},
and then expanding the resulting expression around $q=0$, we can directly compare the
$O(q^0)$ part of the latter with the effective theory susceptibility Eq. \eqref{chi1}.
Performing the comparison order by order in $\delta$ and solving the resulting system for $\tilde{K}_i$
and $v_{Fi}$, $i=0,1$, we obtain for the first order corrections:
\begin{eqnarray}
\tilde{K}_1&=& \tilde{K}(\mu=\infty) (-1+\alpha)^2 \label{K1eq}\; , \\
v_{F1}&=&-v_{F0}(-1+\alpha)(-2+\alpha) \label{vF1eq}\; .
\end{eqnarray}
(the zeroth order terms have already been given in Eq. \eqref{zeroth}).
Eqs. \eqref{leadingK}--\eqref{vF1eq} hold for arbitrary $N_c$ and $N_f$.
In the special case of a single flavor, they can be compared directly to the TBA results,
identifying the bare coupling of 
conformal perturbation theory
and using the mass-coupling relation \cite{Zamolodchikov:1995xk}
\bea
  &&  \left(\frac{\tilde{m}}{4\pi}\right)^2 \rightarrow \mu_{SG}^2=\\
  && \left[m_s\frac{\sqrt{\pi}}{2} \frac{\Gamma(\frac{1}{2}+\frac{b^2}{2-2b^2})}{\Gamma(\frac{b^2}{2-2b^2})}\right]^{4-4b^2}\left(\frac{\Gamma(b^2)}{\pi\Gamma(1-b^2)}\right)^2\; .\nonumber
\eea
where $b^2=1/(2N_c)$.
This comparison is given in Fig. (\ref{Spec}) for the Luttinger parameter,
and in Fig. (\ref{fig_vF}) for the Fermi velocity.
The behavior of $\tilde{K}$ for $N_f>1$ is shown in Fig. (\ref{FigPertSpec}).

\subsection{Baryonic correlation functions at finite density for $N_c =3,N_f =2$}  \label{Strangemetal32}

In this final part of this section, we write down the elementary correlation functions of some of the baryons for the particular case of $N_c=3, N_f=2$.
The first step is to identify the operators for the baryons.  Following the same logic as Ref. \cite{Azaria:2016mqb} these operators involve both the boson $\phi$ of the low energy effective Luttinger theory and operators from the $SU(2)_3$ WZNW model.  For example, the nucleon
of spin-$1/2$ and the $\Delta$-baryon of spin-$3/2$ have the following form in the color singlet sector:
\begin{eqnarray}\label{bar_ops}
n_{R}^{\alpha\beta\gamma} &=& \epsilon^{abc}R_{a\alpha}R_{b\beta}L_{c\gamma} \sim\cr\cr
&& \exp[i\sqrt{2\pi/3}(2\varphi - \bar\varphi)]{\cal F}^{(1)}_{2/5}\bar{\cal F}^{(1/2)}_{3/20} , \cr\cr
\Delta_{R}^{\alpha\beta\gamma} &=&
\epsilon^{abc}R_{a\alpha}R_{b\beta}R_{c\gamma} \sim \exp(3i\sqrt{2\pi/3}\varphi) {\cal F}^{(3/2)}_{3/4}. \cr
&&
\end{eqnarray}
Here ${\cal F}^S_{h}$ are primary fields from the SU$(2)_3$ WZNW model of spin $S$ and scaling dimension $h$.
Both $n_R$ and $\Delta_R$ are right moving - their left moving counterparts are just given by exchanging
$R_{a \alpha} \leftrightarrow L_{a \alpha}$.
The scaling dimensions of the operators in the U(1) and flavor sector for the nucleon are \cite{Azaria:2016mqb}:
\bea
&& (h, \bar h)_c = \Big(\frac{1}{48}(3\sqrt{\tilde{K}} + 1/\sqrt{\tilde{K}})^2, \frac{1}{48}(3\sqrt{\tilde{K}} - 1/\sqrt{\tilde{K}})^2\Big), \nonumber\\
&& (h,\bar h)_f  = (2/5,3/20),
\eea
while that of the $\Delta$-particle is
\bea
&& (h, \bar h)_c = \Big(\frac{3}{16}(\sqrt{\tilde{K}} +1/\sqrt{\tilde{K}})^2,  \frac{3}{16}(\sqrt{\tilde{K}} -1/\sqrt{\tilde{K}})^2\Big), \nonumber\\
&& (h,\bar h)_f =(3/4,0).
\eea
One can see that the scaling dimensions of the fields are determined by the Luttinger parameter, $\tilde K(\mu)$, of the low-energy bosonic theory.

The  correlators for the baryons are like those for Luttinger liquid with spin (i.e. they involve two velocities).  For example, the correlators for right-moving baryons are given by (see Eq. (V.9) of Ref. \cite{Azaria:2016mqb}):
\begin{widetext}
 \bea
 && \la\la n_R(\tau,x)n_R^\dagger(0,0)\ra\ra = Z_n\re^{i k_Fx}\Big[\frac{(\tau v_F +i x)(\tau v_{fl} +i x)}{(\tau v_F -i x)(\tau v_{fl} -i x)}\Big]^{1/4}\Big(\frac{\tau_0^2}{\tau^2 +x^2/v_F^2}\Big)^{\frac{3}{8}\left(\tilde{K}+\frac{1}{9\tilde{K}}\right)}\Big(\frac{\tau_0^2}{\tau^2+x^2/v_{fl}^2}\Big)^{11/20}, \\
 && \la\la \Delta_R(\tau,x)\Delta_R^\dagger(0,0)\ra\ra = Z_{\Delta}\re^{3i k_F x} \Big[\frac{(\tau v_F +i x)(\tau v_{fl} +i x)}{(\tau v_F -i x)(\tau v_{fl} -i x)}\Big]^{3/4}\Big(\frac{\tau_0^2}{\tau^2 +x^2/v_F^2}\Big)^{\frac{3}{8}\left(\tilde{K}+\frac{1}{\tilde{K}}\right)}\Big(\frac{\tau_0^2}{\tau^2 +x^2/v_{fl}^2}\Big)^{3/4} \; ,
 \eea
 \end{widetext}
where $\tau_0 \sim \epsilon_F^{-1}$, while $Z_n$ and $Z_{\Delta}$ are dimensionful constants depending on $\epsilon_F$.
The correlators for the left-moving particles are obtained by replacing $x\rightarrow -x$.

To obtain an idea on how the correlation function appears in frequency-momentum space, we consider the limit $v_F =v_{f l}$ 
In this limit, the retarded Green's function for the nucleons is
 \bea
 && G_n(\omega,q \pm k_F) = \frac{Z}{\omega \mp v_Fq}\Big(\frac{(\omega +i \delta)^2
 -v_F^2q^2}{\epsilon_F^2}\Big)^{\eta}, \nonumber\\
 &&  \eta = -1/5 +\frac{1}{8}\left(\sqrt{3\tilde{K}}+ 1/\sqrt{3\tilde{K}}\right)^2 \; ,
 \eea
 where $\delta$ is an infinitesimal parameter. The general form of the correlation functions of the baryons remains the same for any $N_f$, but the relation of the scaling dimensions to $\tilde{K}$ changes.

We end this section noting that at zero temperature these correlation functions tell us something about instabilities to bosonic condensation in the model.
The way this happens depends on the value of the chemical potential.  An indicator of the instability is provided by the $T\rightarrow0$ behavior of the
susceptibilities corresponding to the creation of the bosons:
\be
\chi=\intop_0^{1/T}d\tau\intop dx\left\langle \hat{T}\mathcal{O}(\tau,x)\mathcal{O}(0,0)\right\rangle \sim T^{-2+2d_\mathcal{O}},
\ee
where $\mathcal{O}$ is the field creating a particular type of boson, and $d_{\mathcal{O}}$ is its scaling dimension.
In case of the scalar meson, $d_{\mathcal{O}}=\frac{1}{6\tilde{K}}+2\Delta$, where $\Delta$ was defined in eq. \eqref{defDelta}.  
The condition for the
instability to occur is $\tilde{K}\geq5/21$. In this case we expect that a meson density wave forms.
Similarly there is an instability to deuteronic superconductivity when $\tilde{K}\leq 7/15$ \cite{Azaria:2016mqb}.

We contrast our results here for quarks transforming in the fundamental with that of adjoint quarks considered in Ref. \cite{Gopakumar_2012}.
For adjoint quarks, there are no color singlet fermions; {\it i.e.}, baryons. 
For fundamental quarks in 1+1 dimensions, there are baryons,
but because their Green functions have cuts not poles, they are incoherent and do not represent well-defined quasi-particles.

\section{From 1+1 to 3+1 dimensions}
\label{sec1d3d}

In QCD, the behavior of the theory in the limit of low and high density is reasonably well understood.

\subsection{Nuclear Matter}

At low densities nuclear matter is strongly interacting, but there is no evidence for non-Fermi liquid behavior.
For example, the shell model provides an excellent description of nuclei, and there the relevant quasi-particles
are certainly nucleons, up to the largest nuclei probed.
This is also expected theoretically in infinite nuclear matter: while pions are massless in the chiral limit, 
they only couple to nucleons through derivative interactions, which do not generate the
infrared divergences required for a non-Fermi liquid \cite{Holt:2014hma}.
Of course there is also superfluidity and superconductivity
for dense nucleons, but that is natural in a Fermi liquid, for the channels in which there is
attraction.

\subsection{Perturbative regime}

In the opposite limit of high density, by asymptotic freedom
perturbation theory in the QCD coupling,
$g$, applies.  The dominant effect is the screening of electric gluons through a Debye mass,
which cuts off any confinement through the exchange of electric gluons.  For quarks
near the Fermi surface, there is also a gap due to color superconductivity,
but this is exponentially small in $1/\sqrt{g^2}$ \cite{Fukushima:2013rx}.
A careful analysis shows that the perturbative regime does
exhibit non-Fermi liquid behavior, but for any reasonable range of coupling, the effects are small
\cite{Brown:2000eh,Boyanovsky:2000bc,Ipp:2003cj,Gerhold:2004tb,Schafer:2004zf,Kitazawa:2005pp,Kitazawa:2005vr,Pal:2011ve,Adhya:2012sq}.

As an approximate rule of thumb, we can assume that the perturbative regime is entered when
the chemical potential, which for on-shell massless quarks is the magnitude of the spatial momentum, is large enough
such that perturbative methods can be used.  In vacuum, certainly perturbation theory,
resummed or not, is not useful below momenta of $1$ GeV.  Thus the perturbative regime cannot set in before
a quark chemical potential of $\mu_{\rm pert} \approx 1$~GeV.  Of course this is a hand waving argument, and could well
be off by a factor of two.  However, heroic computations of the thermodynamic potential to four loop order,
$\sim g^6$, indicate that the this is a reasonable estimate of the quark chemical potential
at which the perturbative regime sets in
\cite{Kurkela:2009gj,Kurkela:2014vha,Kurkela:2016was,Ghisoiu:2016swa,Annala:2017llu,Gorda:2018gpy,Annala:2019puf,Gorda:2021znl,Gorda:2021kme}.

What is not obvious is the behavior of magnetic gluons in the perturbative regime.  At nonzero temperature, and
zero density, this is an old story
\cite{Linde:1980ts,Gross:1980br}.
At very high temperatures, static electric fields are screened by a Debye mass
$m_D \sim \sqrt{g^2} T$.
Loop diagrams bring in factors of $m_D/T$, and so at $T \neq 0$ the perturbative expansion is
a power series not in $g^2$, but in $\sqrt{g^2}$.  This first enters the free energy
at $\sim (\sqrt{g^2})^3$.

Since for bosonic fields
the timelike component of the (Euclidean) momentum at $T \neq 0$ is an integral multiple of $2 \pi T$,
we can naturally divide the theory into non-static momenta, with $p_0 \sim 2 \pi T \neq 0$, and static momenta,
with $p_0 = 0$.  The fermions can then immediately be dropped, as the 
the timelike component of their (Euclidean) momentum is always an
odd multiple of $\pi T$.  This leaves only the interactions of static magnetic fields without quarks,
or a pure gauge theory in three dimensions, QCD$_3$.  The coupling constant for QCD$_3$ 
is $g_3^2 =g^2 T$ \cite{Appelquist:1981vg,Jackiw:1980kv}.
Since this has dimensions of mass, perturbation theory is an expansion in $g^2_3/p = g^2 T/p$, where $p$ is
a characteristic (spatial) momentum of a given diagram.  
Thus perturbative methods are useless to analyse the infrared limit.  Even so, as $g_3^2$ is the only
mass scale in the problem, and as a non-Abelian, super-renormalizable gauge theory,
the theory should confine, with a mass gap proportional to the only mass scale in the problem,
which is $\sim g_3^2$.  This is well confirmed by numerical simulations on the lattice, and
enters into the pressure at $\sim g^6$
\cite{Kajantie:1997tt,Kajantie:2000iz,Kajantie:2002wa,Hietanen:2008tv}.

Now consider the theory at zero temperature and nonzero chemical potential.  First, there is no reason
to separate out static momenta.
As a theory in four dimensions, the timelike component of the momenta, for either
boson or fermion fields, is continuous, and can be as small as the spatial momenta.  There is Debye screening of electric
fields, with a Debye mass $m_D \sim g \mu$.  While this causes some delicacy in computing the free energy
to high order
\cite{Kurkela:2009gj,Kurkela:2014vha,Kurkela:2016was,Ghisoiu:2016swa,Annala:2017llu,Gorda:2018gpy,Annala:2019puf,Gorda:2021znl,Gorda:2021kme},
as in the vacuum, at $\mu \neq 0$  perturbation theory remains an expansion in powers of $g^2$, and {\it not}
in $\sqrt{g^2}$.
Because of the logarithmic infrared divergences in four dimensions,
the powers of $g^2$ are multiplied by powers of logarithms of $\log(m_D^2/\mu^2) \sim \log(g^2)$,
but this is very familiar at zero temperature.

The interesting question is then what happens to {\it magnetic} gluons in cold, dense quark matter?  The effective
theory has three components.  First, there are electric gluons, which are screened by the Debye mass $m_D$.
Second, there are quarks, which (massless or not) develop a small gap about the Fermi surface from
color superconductivity.  Lastly, there are the magnetic gluons, which by gauge invariance
remain massless order by order in perturbation theory.  Magnetic gluons do have logarithmic infrared divergences
perturbatively, but these are very mild, 
$\sim g^2 \log(p^2)$ and $\sim g^2 \log(g^2)$.  Even so, it is {\it in}conceivable that
the magnetic fields remain ungapped: as in vacuum,
surely over large distances a magnetic mass gap is generated {\it non}-perturbatively.

The interesting question is then: what is the {\it ratio} of the mass gap for magnetic gluons
to that for electric gluons?  The former is non-perturbative, while the latter appears at one loop order.
We assume that in all cases, that the ``masses'' are defined in a gauge invariant manner, such as 
from the falloff of two point functions between gauge invariant operators.  

\subsection{Quarkyonic regime}

As the quark chemical potential $\mu$ increases, we then have the following regimes:

\begin{itemize}

\item $\mu_0$: Where a quark chemical potential first matters, determined kinematically from the condition
$\mu_0 = m_B/N_c$, where $m_B$ is the light baryon.

\item $\mu_{\rm Qc} > \mu > \mu_0$: A strongly interacting Fermi liquid of nucleons.

\item $\mu_{\rm pert} > \mu > \mu_{\rm Qc}$:
  The quarkyonic regime, where the free energy is (approximately) that of (interacting)
quarks, but the excitations near the Fermi surface are confined.  

\item $\mu > \mu_{\rm pert}$: The perturbative regime,
  where electric fields are Debye screened and confinement is lost.  Consequently, the value of the Polyakov loop must be
  near unity.  Quarks near the Fermi surface receive a small gap from color superconductivity.  It is a non-Fermi liquid,
  but these effects are mild
\cite{Brown:2000eh,Boyanovsky:2000bc,Ipp:2003cj,Gerhold:2004tb,Schafer:2004zf,Kitazawa:2005pp,Kitazawa:2005vr,Pal:2011ve,Adhya:2012sq}.
Magnetic fields are screened non-perturbatively, with the ratio of that mass scale, to the Debye mass, unclear.

\end{itemize}

We begin by reviewing how the quarkyonic regime can be analyzed by power counting in
(fractional) powers of $N_c$ at large $N_c$ \cite{McLerran:2007qj}.
Then we review recent results from numerical simulations on the lattice for two colors
\cite{Kogut:2001na,Allton:2003vx,Allton:2005gk,Hands:2006ve,Hands:2010gd,Hands:2011ye,Hands:2011hd,Cotter:2012mb,Hands:2012yy,Amato:2015gea,Bornyakov:2017txe,Astrakhantsev:2018uzd,Boz:2019enj,Braguta:2019noz,Iida:2019rah,Wilhelm:2019fvp,Bornyakov:2020kyz,Furusawa:2020qdz,Iida:2020emi,Begun:2021nbf,Bornyakov:2021arl,Bornyakov:2021mfj,Ishiguro:2021yxr}.
By using both methods, we can obtain some guide to the nature of cold, dense QCD, with three colors, and three light
flavors.

Given our analysis in 1+1 dimensions, one of the central questions is where a non-Fermi liquid emerges.
One sign of this is the behavior of the specific heat, which is linear in the temperature as $T \rightarrow 0$,
$C(T) \sim T$, but in a non-Fermi liquid, $C(T) \sim T \log (T)$.  Of course picking out a logarithm from under
a power is always challenging.

Another way to look for the existence of a non-Fermi liquid is to measure the
nature of the quark quasiparticles near the putative Fermi surface.
In a Fermi liquid, the width of the quasiparticle
is much less than the energy, while in a non-Fermi liquid, the width of the quasiparticle is comparable to the energy.
Measuring these quantities in a theory with a local gauge symmetry requires developing gauge-invariant probes.
The simplest way is to add tie two quarks together with a Wilson line \cite{Deryagin:1992rw}, and introduce
\begin{equation}                                                                                                             
G^{ab}(x,y)                                                                                                                
=  \int \overline{q}^{ia}_L(x) \; {\cal P}\exp\left( i g \int^x_y A_\mu(z) dz\right)_{ij} q(y)^{jb}_R \;  ,
\label{chi2}
\end{equation}                                                                                                               
where $q_L$ and $q_R$ are left and right-handed quarks, $i$ and $j$ refer to $SU(N_c)$ color, $a$ and $b$ are
flavor indices, and ${\cal P}$ is path ordering (with the simplest choice a straight line between $x$ and $y$.

These quantities are in principle measureable
through numerical simulations on the lattice, the Functional Renormalization
Group \cite{Fukushima:2021ctq}, {\it etc.}

For three colors, cold, dense QCD can only be studied on the lattice with the quantum computers of the future.
This is not true for two colors, where standard Monte Carlo techniques can be used
\cite{Kogut:2001na,Allton:2003vx,Allton:2005gk,Hands:2006ve,Hands:2010gd,Hands:2011ye,Hands:2011hd,Cotter:2012mb,Hands:2012yy,Amato:2015gea,Braguta:2016cpw,Bornyakov:2017txe,Astrakhantsev:2018uzd,Boz:2019enj,Braguta:2019noz,Iida:2019rah,Wilhelm:2019fvp,Bornyakov:2020kyz,Furusawa:2020qdz,Iida:2020emi,Begun:2021nbf,Bornyakov:2021arl,Bornyakov:2021mfj,Ishiguro:2021yxr}.
However, while the physics of two colors is in some ways very different from that for an odd number of colors,
it is still possible to ask {\it if} a quarkyonic regime arises in the first place.

In all of this we have completely ignored 
the chiral transition.  Unlike the case of zero chemical potential and
nonzero temperature, where chiral symmetry restoration occurs well before deconfinement, the situation is reversed at
low temperature and large chemical potential.  Moving up from the hadronic regime, this is expected to involved the
production of a moat regime \cite{Pisarski:2021qof}, quantum pion liquids \cite{rdp:2020,Pisarski:2020dnx,Pisarski:2021aoz},
a critical endpoint \cite{Bzdak:2019pkr}, and probably many other phenomena, which can be probed experimentally,
both in heavy ion collisions at low energy \cite{Pisarski:2021qof}, and in neutron stars \cite{Pisarski:2021aoz}.

\subsubsection{Large $N_c$}
\label{largeNc}

At least abstractly, the limit of large $N_c$ is especially clean to study \cite{McLerran:2007qj}.
We assume that the number of colors is much larger than the number of flavors, $N_c \gg N_f$, but keep the
factors of $N_f$ to understand the generalization to small $N_c$.

We compare powers of the quark chemical potential $\mu$ versus the renormalization mass scale of QCD,
which for three flavors is $\Lambda_{\overline{MS}} \sim 340$~MeV.  We shall estimate factors of $N_c$
(and $N_f$) assuming that $\Lambda_{\overline{MS}}$ is about the transition temperature for the restoration
of chiral symmetry, which is $T_\chi \approx 154$~MeV.  Thus at the outset, our estimates are only good, at
best, to a factor of two.  This emphasizes the utility of lattice simulations for two colors in Sec.
(\ref{two_colors}).

At zero chemical potential, in the hadronic phase mesons are weakly interacting, with a pressure of order one.
In the deconfined phase, above a temperature $T_d$, the pressure is of order $\sim N_c^2$ from gluons, and
$\sim N_c N_f$ from quarks.  It is natural to expect that the deconfining transition is of first order, with a
latent heat $\sim N_c^2$, and this is confirmed by lattice simulations \cite{Lucini:2012gg,Lucini:2013qja}.
Further, deconfinement is expected to
trigger the restoration of chiral symmetry at a temperature $T_\chi$, since for $N_c \gg N_f$ and $\mu = 0$ the dynamics is
driven by gluons.

The chemical potential doesn't matter until $\mu > \mu_0$, when a Fermi sea of baryons can first form.  Even
though the baryons are strongly interacting, because they are heavy, with a mass $\sim N_c$, the window
in which baryons form is narrow, with $\mu_{\rm Qc} - \mu_0 \sim 1/N_c^2$, Eq. (6) of Ref. \cite{McLerran:2007qj}.

Thus as $\mu$ increases from $\mu_0$, we quickly enter the quarkyonic regime.  Consider first chemical
potentials which are a number of order one times $\mu_0$.  Trivially, the temperature for the
deconfining transition, $T_d(\mu)$, $= T_d(0)$, as when $\mu < \mu_0$ a Fermi sphere
cannot form.  This remains true when $\mu$ is a number of order one times $\mu_0$, since the contribution
of quarks to the free energy is suppressed by a factor of $\sim N_f/N_c$.

This is not true for the restoration of chiral symmetry.
Baryons interact strongly with mesons, and can thereby restore the chiral symmetry.
There is another mechanism which is special to $\mu \neq 0$.
In nuclear matter, the existence of spatially inhomogeneous condensates is familiar,
as pion/kaon condensates
\cite{Overhauser:1960,Migdal:1978az,Kaplan:1986yq,Brown:1995ta}.
For such condensates, a one-dimensional structure forms, and spontaneously breaks both
the rotational and flavor symmetries.  Although
$\langle \overline{q}_L(x) q_R(x) \rangle$ is nonzero locally,
it averages to zero as it winds, periodically, along the direction of the condensate.
To establish a uniformity of notation, we refer to these as pion/kaon chiral spirals.
A pion chiral spiral rotates between, {\it e.g.}, the $\sigma$ and $\pi_3$ directions, with
$\sigma(x)^2 + \pi_3^2(x) = f_\pi^2$.
A more complicated kaon chiral spiral, involving all three flavor directions,
can also emerge at high density \cite{Kaplan:1986yq,Brown:1995ta}.
Even at lower densities, though, a kaon chiral kink can develop; for such a condensate,
$\langle \overline{s}(x) s(x) \rangle$ is nonzero locally, but again averages to zero along the direction
of the kaon chiral kink.

We denote the temperature and chemical potential at which the chiral symmetry is restored as $\mu_\chi(T)$.  In
contrast to Fig. (1) of Ref. \cite{McLerran:2007qj}, though, it is not necessary that $\mu_\chi(T_d) = \mu_0$, up to
corrections of order $\sim 1/N_c^2$; only that $\mu_\chi(T_\chi)$ is within $\mu_0$ by a number of order
one.

In the region where $\mu \sim 1$, although the free energy is quarkyonic, the quasiparticles are uniformly
confined.  Thus it is appropriate to speak of a hadronic chiral spirals, which are best described as a condensate
of mesonic fields.

Now let us consider chemical potentials which grow as a fractional power of $N_c$.
To estimate, we assume for the sake of power counting that in the free energy, $T \sim T_\chi$.  Then
the gluon contribution, $\sim N_c^2 T_\chi^4$, balances the quark contribution, $\sim N_c N_f \mu^4$, when
$\mu_{\rm \pi q} \sim (N_c/N_f)^{1/4} T_\chi$.  For this range, since the contribution of quarks to the free energy
is as large as that of the gluons, we suggest that one goes from a region dominated by hadronic chiral
spirals, to one dominated by quark chiral spirals
\cite{Kojo:2010fe,Pisarski:2018bct,rdp:2020,Pisarski:2020dnx,Pisarski:2021aoz,Pisarski:2021qof}.  We
call them quark, and not quarkyonic, chiral spirals, because both the hadronic and quark chiral spirals occur
in the quarkyonic regime.

Thus there are {\it two} regimes in quarkyonic matter:
\begin{itemize}
\item Hadronic (pion/kaon) chiral spirals from $\mu: \mu_{\rm Qc} \rightarrow \mu_{\rm \pi q} \sim (N_c/N_f)^{1/4} T_\chi$;
\item Quark chiral spirals for $\mu: \mu_{\rm \pi q} \rightarrow \mu_{\rm pert}$.
\end{itemize}

This refines previous analysis in Refs. \cite{McLerran:2007qj,Kojo:2010fe,Pisarski:2018bct}, where the transition
between the two types of chiral spirals was not made precise.  We stress that there is no phase transition between
the region with hadronic chiral spirals and quark chiral spirals; it is just
that in each regime, it is more useful to use one description than the other.

To understand what occurs as $\mu$ increases further, consider
the Debye mass to leading order in perturbation theory:
\begin{equation}
  m_D^2 = g^2 \left( \left( N_c + \frac{N_f}{2} \right) \frac{T^2}{3} + N_f \, \frac{\mu^2}{2 \pi^2} \right) \; .
  \label{debye_mass}
\end{equation}
When $\mu \sim (N_c/N_f)^{1/2} T_\chi$, the quark contribution to the Debye mass is as large as that of the gluons.
This is the perturbative regime, where the screening of electric gluons removes confinement.

It is not evident how to connect this counting powers of $N_c/N_f$ to real scales in QCD.  From
perturbative computations in QCD
\cite{Kurkela:2009gj,Kurkela:2014vha,Kurkela:2016was,Ghisoiu:2016swa,Annala:2017llu,Gorda:2018gpy,Annala:2019puf,Gorda:2021znl,Gorda:2021kme}, for three colors perturbation theory
is valid for $\mu > \mu_{\rm pert} \sim 1$~GeV.  It is difficult to believe that this scale, $\mu_{\rm pert}$,
is very sensitive to either $N_c$ or $N_f$.  In nuclear matter $\mu_0 \sim 300$~MeV.  The large $N_c$
expansion predicts that a quarkyonic regime appears at $\mu_{\rm Qc} = \mu_0 + O(N_f^2/N_c^2)$;
that one goes from a regime of hadronic chiral spirals, to one of quark chiral spirals,
at $\mu_{\rm H-Qc} \sim (N_c/N_f)^{1/4} \mu_0$, and lastly, a 
perturbative regime for $\mu_{\rm pert} \sim (N_c/N_f)^{1/2} T_\chi$.

How does this apply to QCD, where $N_c = N_f = 3$?

\subsubsection{Two colors}
\label{two_colors}

Because the eigenvalues of the quark determinant are complex for more than two colors, numerical simulations of
lattice QCD cannot be carried out with standard Monte Carlo techniques.  This is not true for two colors,
where the quarks lie in a real representation of the gauge group and the eigenvalues of
the quark determinant are real.

Two colors is in some ways rather different from three.  Principally, baryons are composed of two quarks, and so
are bosons.  Thus for two colors, the hadronic regime exhibits Bose-Einstein condensation of the bosonic
baryons, instead of a Fermi sea.  The behavior about the chemical potential where
a Bose-Einstein condensate first forms, $\mu_0 = m_\pi/2$, can be computed in chiral perturbation theory
\cite{Kogut:1999iv,Kogut:2000ek,Son:2000by}.

We summarize some results from lattice studies of QC$_2$D with two flavors.
This includes groups from Russia, Refs.
\cite{Braguta:2016cpw,Bornyakov:2017txe,Astrakhantsev:2018uzd,Braguta:2019noz,Bornyakov:2020kyz,Begun:2021nbf,Bornyakov:2021arl,Bornyakov:2021mfj}
and Japan, Refs. \cite{Iida:2019rah,Iida:2020emi,Furusawa:2020qdz,Ishiguro:2021yxr}.

In Ref. \cite{Braguta:2016cpw}, $m_\pi \sim 380$~MeV, so $\mu_0 \sim 190$~MeV.  For
$\mu: 190 \rightarrow 350$~MeV, nuclear matter is a dilute baryon gas, and the value of the
diquark condensate agrees with chiral perturbation theory.  For higher density, one enters a regime
of dense baryons.

In Refs. \cite{Astrakhantsev:2018uzd} and \cite{Astrakhantsev:2020tdl}, heavier pions were used, with
$m_\pi \sim 740$~MeV, and $\mu_0 = 370$~MeV.  Chiral perturbation theory is useful to
describe the diquark condensate for $\mu : \mu_0 \rightarrow \mu_{\rm Qc}$, where
$\mu_{\rm Qc} \sim 540$~MeV; for $\mu > \mu_{\rm Qc}$, the diquark
density exceeds the value expected from chiral perturbation theory, computed at lowest order.

The renormalized Polyakov loop is a monotonically
increasing function of $\mu$, but even by $\mu \sim 1$~GeV, it is only $\sim 1/3$ its value at the highest
$\mu$ which was probed, $\mu \sim 2$~GeV.  This suggests that $\mu_{\rm pert} \sim 1$~GeV.  This is also confirmed by
measurements of the Debye mass, for which $m_D/\mu$ is approximately constant from $\mu: 1 \rightarrow 2$~GeV,
Figs. (12)-(14) of Ref. \cite{Astrakhantsev:2018uzd}.   From Ref. \cite{Astrakhantsev:2020tdl}, a Fermi sphere forms
above $\mu \sim 900$~MeV, with diquarks forming as expected from color superconductivity.  

(Incidentally, in Fig. (8) of Ref. \cite{Astrakhantsev:2020tdl}, the topological susceptibility varies {\it very}
slowly as $\mu$ increases, and is still $\approx 75\%$ of its value in vacuum when $\mu = \mu_{\rm pert}$.
See, also, Fig. (13) of Ref. \cite{Iida:2019rah}, where qualitatively similar results are found.
This agrees with the analysis of Ref. \cite{Pisarski:2019upw}.  There, the relative factors of $N_c$ and $N_f$
in the expression for the Debye mass in Eq. (\ref{debye_mass}) suggests that the topological susceptibility
varies {\it much} more slowly with $\mu$ than it does with $T$.)

It is worth contrasting the results for two colors with the qualitative estimates for $N_c = \infty$.  First,
for two colors, the regime of nuclear matter is rather broad, as
$\mu_{\rm Qc} \approx 1.5 \mu_0$.  Second,
as indicated by naive expectation, again $\mu_{\rm pert}$ is about $\approx 1$~GeV.  In between lies dense
quark matter, which we can hopefully term quarkyonic.

More detailed questions await further study.
It is not clear that hadronic chiral spirals will win over Bose-Einstein condensation at low
density, or if quark chiral spirals emerge in the quarkyonic phase.
Determining the appearance of quark chiral spirals is challenging,
as is determining if the quarkyonic regime is a non-Fermi liquid.  Both would be useful to study.

One quantity which the lattice has studied is the
ratio of the electric to magnetic masses.  In Ref. \cite{Bornyakov:2020kyz}, the electric
and magnetic propagators were fit to forms suggested by solutions of the Schwinger-Dyson equations.  These gauge
variant results find that the electric mass is uniformly heavier than the magnetic.   This does not coincide with
our expectation that the quarkyonic regime is confining.

Building upon previous analysis in Refs. \cite{Iida:2019rah,Iida:2020emi,Furusawa:2020qdz}, however,
Ref. \cite{Ishiguro:2021yxr} studied the profile of the color flux tube.  They find that even in high density
superfluid phase, which we term quarkyonic, that the color electric field {\it is} squeezed into a flux
tube, as it is in the hadronic phase, for chemical potentials as large as $\mu \sim 840$~MeV.
\footnote{ E. Itou, private communication.}
This, along with the measurements of the renormalized Polyakov loop,
confirms that it is reasonable to assume that electric fields {\it are} confined in the quarkyonic phase,
for $\mu_{\rm Qc} < \mu < \mu_{\rm pert}$.

Thus simulations for $N_c = N_f = 2$ are essential to demonstrate that the estimates for large $N_c$,
with $N_c \gg N_f$, are a reasonable guide to QCD, where $N_c = N_f = 3$.

\subsection{Quark chiral spirals}

To describe quark chiral spirals, we start 
with the Gribov-Zwanziger form of the electric propagator,
\begin{equation}
  \Delta_{00}(p) = \frac{\sigma}{(p^2)^2} \; .
\end{equation}
Assuming that we can integate over the two spatially transverse directions, gives a gluon
propagator in 1+1 dimensions,
\begin{equation}
  \int d^2p_\perp \frac{\sigma}{(p_0^2 + p_z^2 + p_\perp^2)^2} =  \frac{\sigma}{p_0^2 + p_z^2} \; .
\end{equation}
We shall use this expression to give a qualitative estimate of various quantities below.
We note that explicit solutions of the Schwinger-Dyson and Functional Renormalization Group equations
find a more complicated form for the gluon propagator \cite{Fukushima:2021ctq}.
Nevertheless, this expression is useful to give a rough estimate.
Assume that one is in the regime of quark chiral spirals, $\mu: \mu_{\pi q} \rightarrow \mu_{\rm pert}$.
Following Refs. \cite{Kojo:2010fe,Pisarski:2018bct},
the confinement scale in 3+1 dimensions is proportional to the string tension,
$\sim \s$.  If $\mu \gg \s \sim \mu_0$, then it
is natural that the Fermi sphere breaks up into a series of patches, with a weak coupling between
different patches. The condensate of density waves are described by vectors ${\bf Q}_a$, 
where the Fermi surface is then covered by these patches, with $a=1,2\ldots{\cal N}$.
In momentum space at the edge of the Fermi sphere, the
size of each patch is $\sim \s/v_F$, where $v_F$ is the Fermi velocity.  Thus the number of
patches is ${\cal N} \sim k_F^2/\s$.

\begin{figure}[!htb]
\includegraphics[width=\linewidth,clip]{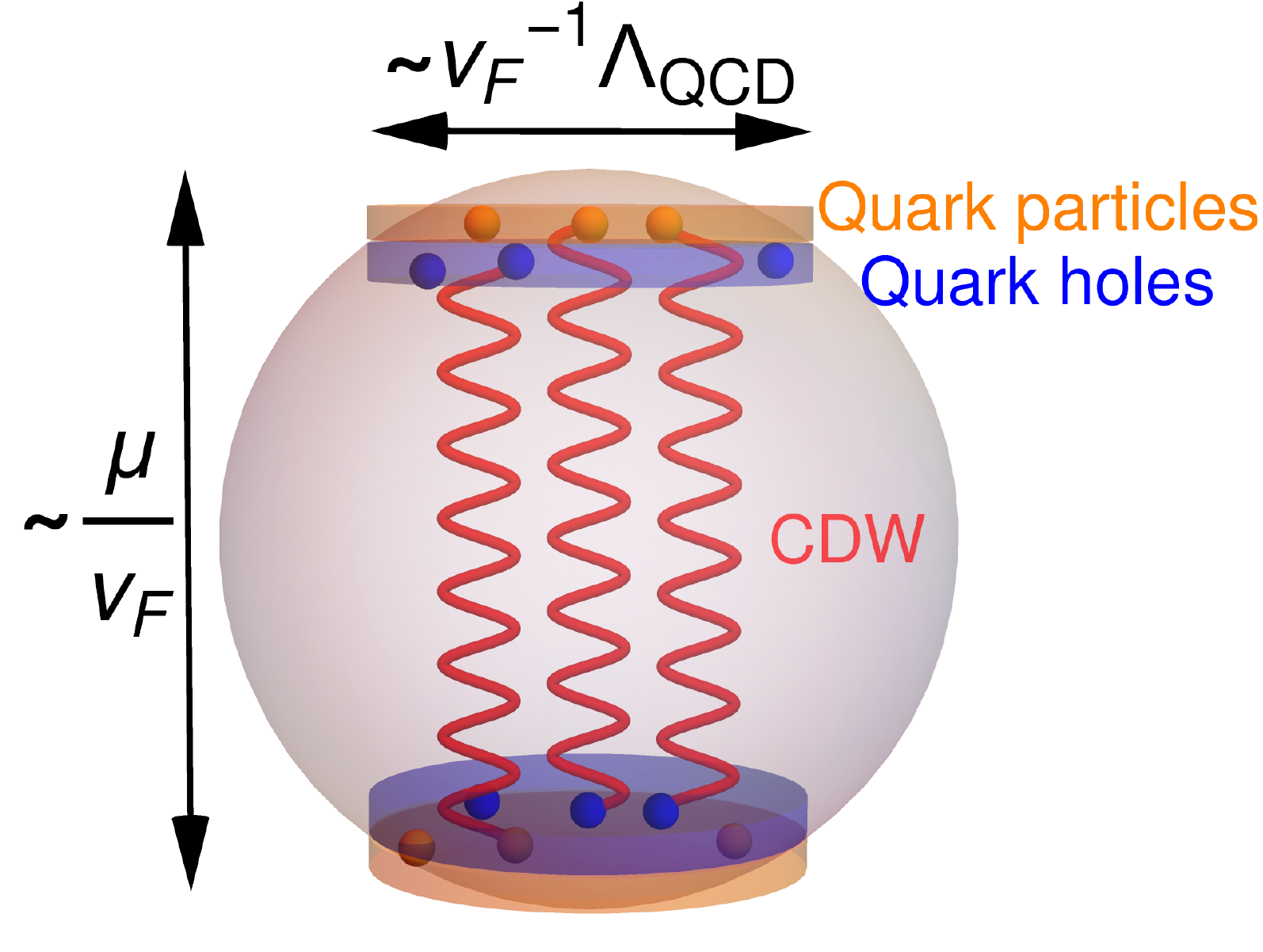}
\caption{Due to the singular character of the  interaction,
  quarks at the Fermi surface scatter back and forth inside patches of a particular size. 
 \label{Patches} }
\end{figure}

In the regime of quark chiral spirals, $k_F \sim \mu \geq (N_c/N_f)^{1/4} \mu_0$, and so the number of patches is large,
${\cal N} \sim (N_c/N_f)^{1/2}$.
The dynamics in a single patch is illustrated in Fig. (\ref{Patches}), where for the purposes of illustration
we have exaggerated the size of the patch, as the figure only applies to the regime 
where there are {\it many} patches, ${\cal N} \gg 1$.
In Eqs. (11) and (17-19) of Ref. \cite{Pisarski:2018bct}, 
the corresponding slow order parameter fields for the density waves
are quasi-one-dimensional  modifications of those for QCD in 1+1 dimensions,
as described in the previous sections. Below we elaborate on the effects of these modifications.

\begin{figure}[!htb]
\includegraphics[width=\linewidth,clip]{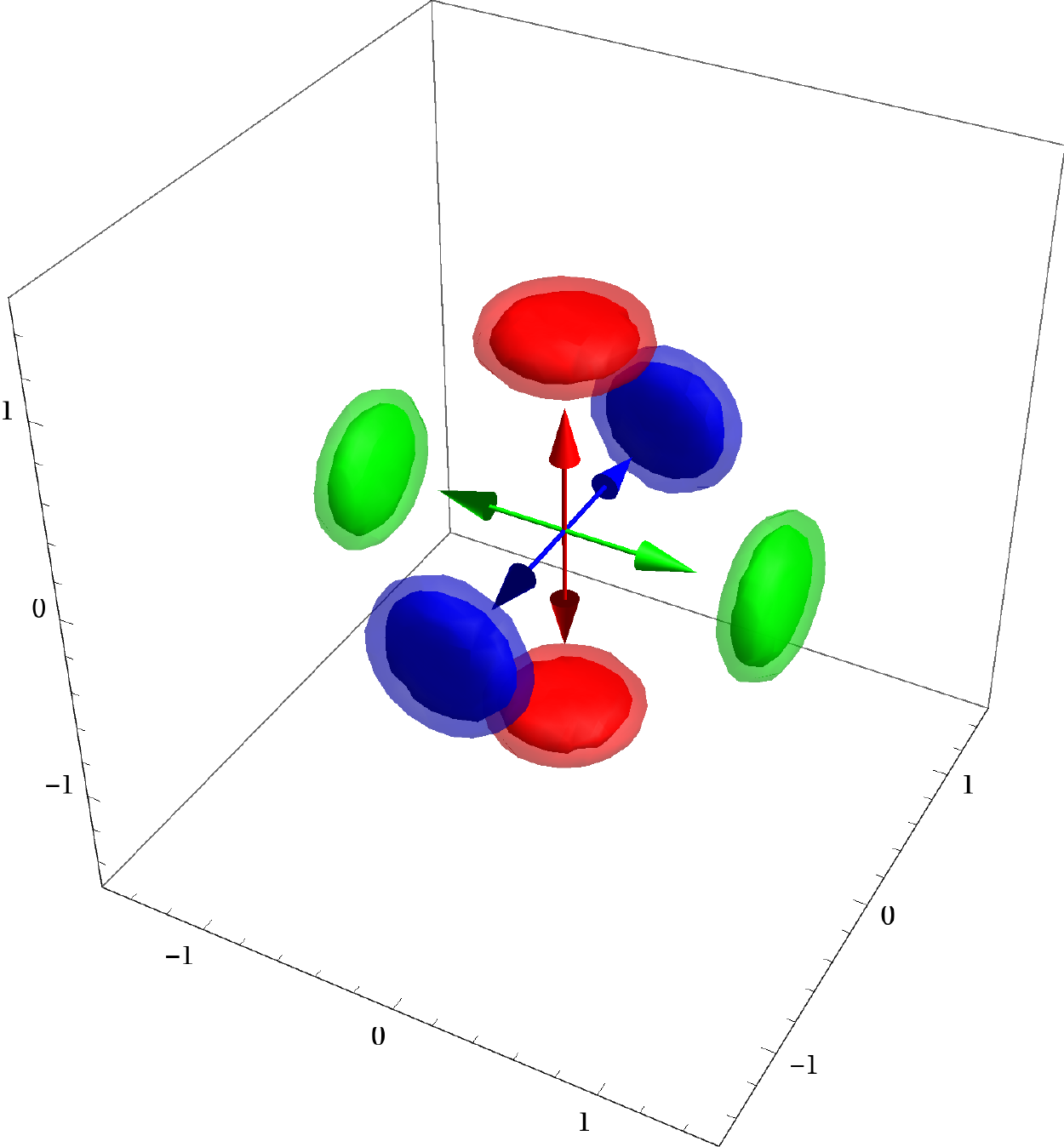}
\caption{ The contour plots of the meson spectrum Eq. (\ref{meson})  for the smallest
  possible number of patches, six, for a cubic arrangement of density waves.  We
  set $\alpha =1$, with the axes wave vectors $q_{x,y,z}/2k_F$. 
 \label{FigMeson} }
\end{figure}

The case of three pairs of patches is shown in Fig. (\ref{FigMeson}), and illustrates what happens with
hadronic (pion/kaon) chiral spirals, in the range $\mu: \mu_{Qc} \rightarrow \mu_{\pi q}$.  

Each pair of patches has its own set of fields;  a scalar $U(1)$ and a matrix $SU(2N_f)$ field $G_{-{\bf Q}} = G^+_{\bf Q}$.
To distinguish them we have to assign to each field a subscript ${\bf Q}$, with the $2 N_f$ for
an enlarged spin-flavor symmetry.
Since wave vectors of the density waves in different patches are not parallel to one other,
the direct coupling of the $G_{\bf Q}$ complex matrix fields  with different ${\bf Q}$ is suppressed, at
least at large ${\cal N}$.
However, inside a given patch the excitations have a transverse dispersion which makes the case in 3+1 dimensions
somewhat different from that in 1+1 dimensions.
The energies of the meson fields depend upon the momenta transverse to the wave vector of their
patch ${\bf Q}_a$, where $|{\bf Q}_a |\approx 2k_F$. 
As for smectic liquid crystals, the stiffness in the direction transverse to ${\bf Q}$ vanishes,
and the dispersion relation is anisotropic.
The action for each pair is the $U(1)\times SU_{N_c}(2N_f)$ WZNW model,
augmented by a term which depends upon the transverse momenta:
\bea
-\; \frac{\alpha v_F}{(2k_F)^{-4}}\; \mbox{Tr}\Big(G^+_{\bf Q}[{\bf Q}\times \vec\nabla]^2G_{\bf Q}\Big)^2 \; ,
\eea
where $\alpha$ is a numerical coefficient. Linearizing the fluctuations of $G$, the mesonic modes
have the dispersion relation,
   \bea
   && \omega_{\bf Q}({\bf q})^2 = \label{meson}\\
   && (v_F/2k_F)^2\Big\{[{\bf Q}\cdot({\bf q} - {\bf Q})]^2 + \alpha [({\bf Q}\times{\bf q})^2]^2/4k_F^2\Big\}.\nonumber
   \eea
The transverse fluctuations introduce a new energy scale, as
the width of the fluctuations in the transverse direction, $W \sim \s/k_F\sim k_F/{\cal N}$. 
This scale marks a crossover from a regime in 1+1 dimensions to one where 
where transverse fluctuations are important.
The separation of $U(1)$ charge - and spin-flavor  holds above the crossover scale $W$, where  the  composite
description of Eqs. (\ref{bar_ops}) apply.
The baryons are incoherent for energies above $W$, and coherent for those below.
Below $W$, the transverse fluctuations enter, and
one cannot directly apply the bosonization formulae
to calculate baryon correlation functions.
A reasonable  assumption is that since the spectrum of the
fundamental excitations, which are mesons,
is gapped everywhere except at the discrete set of points in momentum space, the baryons lack an extended Fermi surface.
   
The vanishing of the Fermi surface affects the thermodynamics for temperatures $T \sim W$.
At temperatures above $\sim W$
the system is a Luttinger liquid, with the specific heat linear in temperature $\sim T/\epsilon_F$.

\begin{figure}[!htb]
\includegraphics[width=\linewidth,clip]{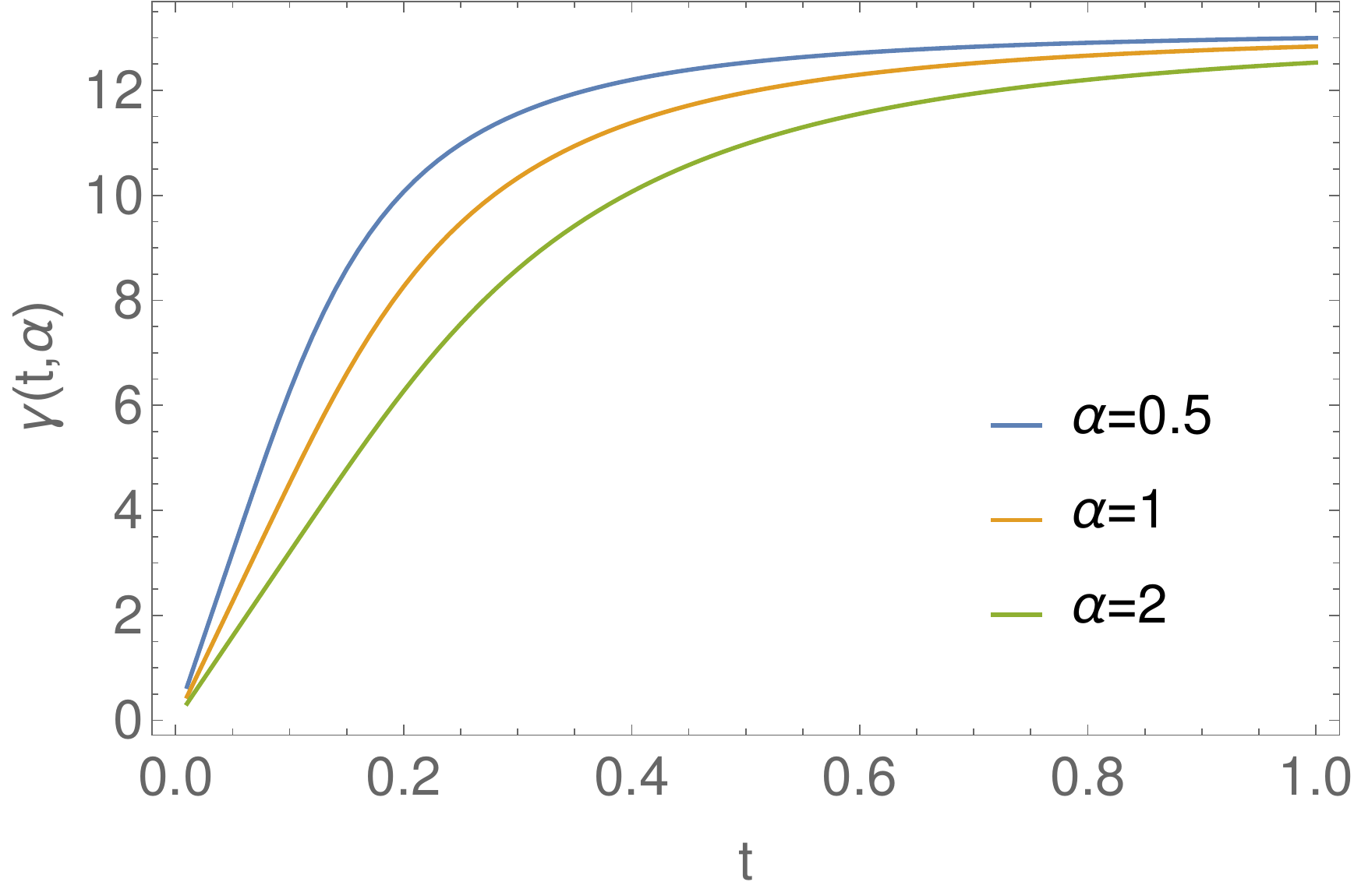}
\caption{The specific heat divided by temperature, $\gamma(t,\alpha)$ in Eq.(\ref{gamma}), for $\alpha = 0,5,1,2$, as a function
  of $t = T/W$. }
 \label{heat} 
\end{figure}

At temperatures below $\sim W$, the specific heat is $\sim T^2({\cal N}/\epsilon_F)$.
To get a more detailed description of the crossover we treat the size of the patch as a
rigid cut-off for the transverse momentum, and normalize behavior for $T \gg W$
to the asymptotic form for the specific heat for a conformal field theory.  In this case,
it is for a $U(1)\times SU_{N_c}(2N_f)$ WZNW model, where the central charge is
$C = 1+ N_c(4N_F^2-1)/(2N_f +N_c)$. Then the temperature dependence of the specific heat is given by the formula
\bea
&& C_v/T = 2k_F^3 C  \gamma(T/W;\alpha),  \nonumber\\
&& \gamma(t,\alpha) = t\int_0^{\infty} dx \int_0^{1/t} dy \frac{(x^2+ \alpha y^2)}{\sinh^2(\sqrt{x^2 +\alpha y^2}/2)},\label{gamma}
\eea
where the function $F(t,1)$ is depicted in Fig. (\ref{heat}).   

The softness of the transverse dispersion  strengthens the thermal fluctuations preventing a development of any kind of long range order  at finite temperature. NFL is a critical state and is incompatible with long range order. Formally the destruction of the long range order is related to the divergence of the integral,
\bea
T\sum_n \int \frac{\rd q_{\parallel}\rd^2 q_{\perp}}{(2\pi Tn)^2 +  \omega^2(q)} =\int \frac{\rd q_{\parallel}\rd^2 q_{\perp}}{2\omega(q)}\coth[\omega(q)/2T],\nonumber
\eea
measuring the relative fluctuations of the order parameter fields. This integral converges at $T=0$ and logarithmically diverges at nonzero temperature when it is dominated by the $n=0$ Matsubara frequency: 
 \bea
 T \int \frac{\rd q_{\parallel}\rd^2 q_{\perp}}{q^2_{\parallel} + \alpha q^4_{\perp}/4k_F^2} .
\eea
In
Ref. \cite{Pisarski:2018bct} we argued that at finite $T$ the Abelian sector of the
theory remains critical with temperature dependent power law correlations,
but the non-Abelian flavor fluctuations acquire a correlation length exponentially large in $1/T$.
This difference is related to the fact that the low energy theory for the Abelian sector is Gaussian,
but the non-Abelian sector is described by
the nonlinear sigma model where the interactions generate a finite correlation length.

We comment that if the condensate arises from a moat spectrum, then the correlation lengths of the non-Abelian fields
remains finite at zero temperature \cite{Pisarski:2021aoz}.  This correlation length is exponentially large
in the coupling constant, and implicitly we assume that this correlation length is larger than all scales above.
Even if this scale is commensurate with those above, the results do not change qualitatively.  

\subsection{Neutrino emission}

Most properties of cold quark matter are difficult to observe directly.  Because
neutron stars are transparent to neutrinos, neutrino emission can be inferred, albeit
indirectly, by the cooling of neutron stars
\cite{Yakovlev:2000jp,Yakovlev:2004iq,Page:2004fy,Page:2010aw,Shternin:2010qi,Potekhin:2015qsa,Schmitt:2017efp}.

In pion/kaon condensates, it is known that effective quasiparticles enhance neutrino emission, through
their decay into a quasiparticle neutron-proton pair
\cite{Yakovlev:2000jp,Yakovlev:2004iq,Page:2004fy,Page:2010aw,Shternin:2010qi,Potekhin:2015qsa}.
As discussed in the previous section, there are strong similarities between pion/kaon condensates and
the condensates in the quarkyonic phase.  

Thus it is natural to expect that neutrino emission is similarly enhanced in the quarkyonic phase \cite{Tsvelik:2021ccp}.  
Fluctuations of the WZNW matrix fields decay into a virtual nucleon pair,
and thereby through the weak interaction into a lepton neutrino pair \cite{Iwamoto:1982zz}:
\bea
 && {\cal L}_{W} \sim  \ri g_W \sum_{\bf Q} [\bar e({\bf Q} \cdot {\vec\gamma})(1-\gamma_5)\nu_e + 
  \bar \mu({\bf Q} \cdot {\vec \gamma})(1-\gamma_5)v_{\mu}] \times\nonumber\\
 && \Big(\mbox{Tr}(G^+ \; {\bf Q} \cdot {\vec\nabla} G\hat\tau^+)\Big) +H.c.,\label{W}
 \eea
where $G$ is the SU(2) flavor WZNW matrix field for a given patch vector ${\bf Q}$,
$\vec\gamma$ are Dirac gamma matrices with space-like indices, and
$\hat\tau^+$ is the Pauli matrix acting on flavor indices of $G$ \cite{Pisarski:2018bct}
(we dropped the subscript ${\bf Q}$).

The neutrino emissivity is
\bea
{\cal E}_{\nu} = 2\int \frac{\rd^3 p_{\nu}}{(2\pi)^3} p_{\nu} \frac{\p }{\p t}f_{\nu}(t,{\bf p}),
\eea
where $f_{\nu}$ is the neutrino distribution function. Following Ref. \cite{Schmitt:2017efp},
\bea
{\cal E}_{\nu} \sim \mu_e \rho(\epsilon_F) T^6 g^2_s(T),
\eea
where $\mu_e$ is the chemical potential of the electrons,
$\rho(\epsilon_F)$ is the quark density of states and
$g_s(T) $ is the running coupling constant. Since $\rho(\epsilon_F)$ is proportional to the specific heat, we can write
\bea
{\cal E}_{\nu} \sim \mu_e (\frac{\p S}{\p T}) T^6 g^2_s(T),
\eea
where $S$ is the entropy.
Substituting this into the equation $\frac{\p F}{\p t} = - {\cal E}_{\nu}(T)$,
where $F$ is the free energy per unit volume, we obtain the equation for the temperature as a function of time:
\bea
\frac{\p T}{\p t} = - A\mu_e (\frac{\p \ln S}{\p T}) T^6 g^2_s(T), \label{rate}
\eea
where $A$ is a temperature independent proportionality coefficient. 

For $\epsilon_F > T > W \sim \epsilon_F/{\cal N}$, the density of
states is constant, while at lower temperatures we replace
$\rho(\epsilon_F)$ by $T/W$. The right hand side of Eq. (\ref{rate})
contains $\ln S\sim n\ln T$, where $n=1$ for $T>>W$ and $n=2$ for $T<<W$ .
The temperature dependence of the strong coupling constant $g_s(T)$ can be neglected, as that is
only logarithmic. So approximately,
\bea
T = \frac{T_0}{(1 + t/\tau)^{1/4}}, ~~ 1/\tau \sim n \, T_0^4g^2(T_0)\mu_e.
\eea
where $T_0$ is the  temperature at time $t=0$. The same time dependence is found with weakly interacting
quarks, with the only difference the overall numerical coefficient in front.  This difference, however,
is important, and it would be well worth studying in detail.

\section{Conclusions}

 In this paper we considered the strange metal phase of QCD in 1+1 dimensions, and
 explored the consequences for quarkyonic matter in 3+1 dimensions. In 1+1 dimensions a
 strange metal exists above a given quark density; it is a  critical
 phase,  where nucleon excitations are incoherent and essentially a
 Tomonaga-Luttinger liquid. In 3+1 dimensions the low energy theory remains
 very similar, as it remains a nonlinear sigma model for  the collective mesonic
 modes, with the baryons topological excitations of the meson
 field. The spectrum of the mesons is highly anisotropic, and is very
 soft in the direction tangential to the Fermi surface. This
 introduces an extra energy scale related to the crossover from the
 regime of one dimensional fluctuations at high energy, to the
 regime of anisotropic three dimensional fluctuations at low
 energies. Above this energy scale the Luttinger liquid description of
 the quarkyonic state remains valid  and the nucleons are incoherent.
 We have not been able to determine whether fermionic quasiparticles
 exist at lowest energies. 
  We also explored the consequences for neutrino emission from neutron stars.

\acknowledgements
ML, AMT and RMK are supported by Office of Basic Energy Sciences,
Material Sciences and Engineering Division, U.S. Department of Energy (DOE) under Contract No. DE-SC0012704.
RDP was supported by the U.S. Department of Energy under contract DE-SC001270,
by B.N.L. under the Lab Directed Research and Development program 18-036, and by
the U.S. Department of Energy, Office of Science, National Quantum Information Science Research Centers,
Co-design Center for Quantum Advantage (C$^2$QA) under contract DE-SC001270.
We thank Toru Kojo for allowing us to use a figure of his, which we modified into our Fig. (5).
RDP also thanks Vitaly Bornyakov, Victor Braguta, Etsuko Itou, and Roman Rogalyov for
helpful discussions of their work.

\appendix
 \section{Facts about Quantum Ising model}
\label{app_ising}
The quantum Ising model on a lattice is defined as 
\bea
H = -\sum_n\Big(J\s^x_n\s^x_{n+1} - h\s^z_n\Big), \label{Ising1}
\eea
where $\s^a$ are the Pauli matrices. It allows a self-dual representation in terms of other Pauli matrix operators defined on the dual lattice:
\bea
\mu^z_{n+1/2} = \s^x_n\s^x_{n+1}, ~~ \mu^x_{n+1/2} = \prod_{j=1}^{n}\s^z_j, \label{Ising2}
\eea
so that
\bea
H = - \sum_n\Big(J\mu^z_{n+1/2} - h\mu^x_{n+1/2}\mu^x_{n-1/2}\Big).
\eea
It is obvious that at $J>>h$ the ground state of (\ref{Ising1}) is such that $\la\s^x\ra \neq 0$ and at $J << h$ the ground state of (\ref{Ising2}) is such that $\la\mu^x\ra \neq 0$. Since $\s^x$ and $\mu^x$ cannot order simultaneously, we conclude that the point $J=h$ is critical separating two phases of model (\ref{Ising1}). 

Using the Jordan-Wigner transformation, one can transform
(\ref{Ising1}) into a Hamiltonian of noninteracting Majorana fermions
$\rho_n,\eta_n$ with anticommutation relations 
\begin{equation}
\rho_n,\rho_m\} = \{\eta_n,\eta_m\} = \delta_{nm}, ~~\{\rho_n,\eta_m\}=0,
\end{equation}
i.e., 
\bea
H = i\sum_n \Big( J\rho_n\eta_{n+1} -  h\rho_n\eta_n\Big).
\eea
As far as applications to quantum field theory are concerned, we need to pass to the continuum limit via defining $n = x/a_0$ with $a_0\rightarrow 0$ the lattice spacing or, alternatively, momentum cutoff. The continuum limit of $\s^x$ and $\mu^x$ operators become the order and disorder fields $\s(x)$ and $\mu(x)$.  The continuum limit of the Majoranas is related to the chiral fields used in the main text as
\bea
\frac{1}{a_0}\rho_n  \rightarrow \chi_R +\chi_L, ~~ \frac{1}{a_0}\eta_n \rightarrow \chi_R -\chi_L.
\eea
At the critical point, $J=h$, the fields $\s,\mu$ are primary fields with  conformal dimensions $(1/16,1/16)$.  The right- and left-moving Majorana fields have conformal dimensions $(1/2,0)$ and $(0,1/2)$ respectively.
 
Since two species of real Majorana fermions form a complex Dirac fermion which in turn can be bosonized and described by the Gaussian model of a bosonic field $\Phi$, one can establish a correspondence between $\s,\mu$ and the bosonic field. This correspondence was first given by Itzykson and Zuber as \cite{Zuber:1976aa} 
\bea
&& \mu_1\mu_2 \sim \cos(\sqrt\pi\Phi), ~~ \s_1\s_2 \sim \sin(\sqrt\pi\Phi), \nonumber\\
&& \s_1\mu_2 \sim \cos(\sqrt\pi\Theta), ~~ \mu_1\s_2 \sim \sin(\sqrt\pi\Theta),
\eea
where $\Theta =\varphi -\bar\varphi$ is a field dual to $\Phi$. These are the identities that we use in the main text. 

As is shown in the main text, the color and flavor sectors in the model with $N_c=N_f =2$ are described by six Majorana fermions which naturally split into two groups of three. The interaction of the color currents include three Majorana fermions corresponding to three critical Ising models. The interaction leads to a spontaneous breaking of the $[Z_2]^3$ symmetry in the color sector generating vacuum averages either for $\mu$ or $\s$ fields of the corresponding Ising models.  

\section{Thermodynamic Bethe Ansatz}
\label{app_tba}
In the sine-Gordon theory, on-shell scattering is purely elastic (in
the sense that the number of particles and the set of rapidities always
has to be the same in the in and out state) and it factorises completely.
Knowledge of the $2\rightarrow2$ elastic scattering phases between
all particle types provides a complete description of all scattering
processes. This makes it an integrable model. At a general sine-Gordon
coupling, scattering is not completely diagonal due to the possibility
of $s\bar{s}\rightarrow\bar{s}s$ scattering. In our cases of interest
($N_{c}\in\mathbb{Z}$) these off-diagonal amplitudes are exactly
zero, which simplifies the discussion.

To derive the thermodynamics, it is useful to adapt a first quantized
viewpoint, where a state is described by a multiparticle wavefunction.
For a moment let us restrict the model onto a circle of circumference
$L$. Since it is a relativistic system, interactions are local. Particles
propagate freely between pointlike interactions, when they acquire
a phase determined by the S-matrix element. This provides the quantization
condition for the particles up to exponentially small corrections
for large $L$:
\be
e^{ip_{j}L}\prod_{k\ne j}S_{r_{j}r_{k}}(\theta_{j}-\theta_{k})=1,
\ee
where $r_{i}$ specifies the type of particle $i$. Taking the logarithm,
one has
\begin{equation}
m_{r_{j}}L\sinh\theta_{j}+\sum_{k\neq j}\delta_{r_{j}r_{k}}(\theta_{j}-\theta_{k})=2\pi n_{j},\quad n_{j}\in\mathbb{Z},\label{eq:BetheAnsatzGen}
\end{equation}
where
\be
\delta_{r_{j}r_{k}}\left(\theta\right)=-i\log S_{r_{j}r_{k}}\left(\theta\right).
\ee
The set of equations (\ref{eq:BetheAnsatzGen}) is commonly referred
to as the Asymptotic Bethe Ansatz or Bethe-Yang equations. The energy
and momentum of multiparticle states are simply given as a sum of
one-particle states with the obtained rapidities
\be
E=\sum_{j}m_{r_{j}}\cosh\theta_{j},\quad P=\sum_{j}m_{r_{j}}\sinh\theta_{j}.
\ee
In the sine-Gordon model these $S$-matrix elements are well-known.
They have the form ($N_{c}\in\mathbb{Z}$)
\bea
&S_{ss}\left(\theta\right) = S_{s\bar{s}}\left(\theta\right)=S_{\bar{s}s}\left(\theta\right)=S_{\bar{s}\bar{s}}\left(\theta\right)\\
&= -e^{\intop_0^\infty \frac{dt}{t}\frac{\sinh(\nu-2)t}{\sinh(t)\cosh((\nu-1)t)}\sinh\left(\frac{2t\theta(\nu-1)}{i\pi}\right)}
\eea

\bea
&S_{sn}\left(\theta\right)  =S_{ns}\left(\theta\right)=S_{\bar{s}n}\left(\theta\right)=S_{n\bar{s}}\left(\theta\right)\\
&  =\frac{\sinh\theta+i\cos\frac{n\pi}{2\left(\nu-1\right)}}{\sinh\theta-i\cos\frac{n\pi}{2\left(\nu-1\right)}}\prod_{k=1}^{n-1}\frac{\sin^{2}\left(\frac{n-2k}{4\left(\nu-1\right)}\pi-\frac{\pi}{4}+i\frac{\theta}{2}\right)}{\sin^{2}\left(\frac{n-2k}{4\left(\nu-1\right)}\pi-\frac{\pi}{4}-i\frac{\theta}{2}\right)}
\eea

We now focus on the case of finite soliton density. In the ground
state of our setup, all soliton states are filled up to a treshold
rapidity $B$, and none above. To obtain the continuum limit of the
quantization condition, we introduce the rapidity density $\rho_s\left(\theta\right)$.
Denoting the total number of \textit{soliton} one-particle states between
rapidity $0$ and $\theta$ with $N\left(\theta\right)$, the rapidity
density is defined as $\rho_s\left(\theta\right)=L^{-1}\frac{dN}{d\theta}$.
Therefore, $n_{j}=L\intop_{0}^{\theta_{j}}\rho_s\left(\theta^{\prime}\right)d\theta^{\prime}$.

Let us first consider the finite-density ground state. The sum in
eq. (\ref{eq:BetheAnsatzGen}) can be rewritten as an integral over
rapidities (dropping the index $j$)
\be
m_{s}L\sinh\theta+L\intop_{-B}^{B}d\theta^{\prime}\rho_s\left(\theta^{\prime}\right)\delta_{ss}\left(\theta-\theta^{\prime}\right)=2\pi L\intop_{0}^{\theta}\rho_s\left(\theta^{\prime}\right)d\theta^{\prime}
\ee
Differentiation with respect to $\theta$, we obtain an integral equation
for $\rho_s\left(\theta\right)$,
\be
\frac{m_{s}}{2\pi}\cosh\theta=\rho_s\left(\theta\right)+\intop_{-B}^{B}d\theta^{\prime}\rho_s\left(\theta^{\prime}\right)\mathcal{K}_{ss}\left(\theta-\theta^{\prime}\right)
\ee
where $\mathcal{K}_{ss}\left(\theta\right)=-\frac{1}{2\pi}\partial_{\theta}\delta_{ss}\left(\theta\right)$.
The solution can be written formally as
\be
\rho_s\left(\theta\right)=\left(1-\mathcal{K}_{ss}\right)^{-1}\left[\frac{m_{s}}{2\pi}\cosh\theta\right]
\ee
The energy of the finite-$B$ ground state is thus
\be
E_{0}=\sum_{j}m_{s}\cosh\theta_{j}=L\intop_{-B}^{B}\rho_s\left(\theta\right)\left(m_{s}\cosh\theta-\mu\right)d\theta
\ee
The operator $\left(1-\mathcal{K}_{ss}\right)^{-1}$ is symmetric,
so the ground state energy is equivalently written as
\be
E_{0}=Lm_{s}\intop_{-B}^{B}\frac{d\theta}{2\pi}\cosh\theta\cdot\epsilon_{s}\left(\theta\right)
\ee
where the soliton spectral gap $\epsilon_{s}\left(\theta\right)$
is the solution of the Fredholm equation with inhomogenity $m_{s}\cosh\theta-\mu$,
\be
\epsilon_{s}\left(\theta\right)=\left(1-\mathcal{K}_{ss}\right)^{-1}\left[m_{s}\cosh\theta-\mu\right]
\ee
and the total $U(1)$ charge is the number of solitons in the condensate,
\be
Q_{0}=L\intop_{-B}^{B}\rho_s\left(\theta\right)d\theta=Lm_{s}\intop_{-B}^{B}\frac{d\theta}{2\pi}\cosh\theta\cdot\zeta\left(\theta\right)
\ee
where $\zeta\left(\theta\right)=\left(1-\mathcal{K}_{ss}\right)^{-1}\left[1\right]$
is the dressed charge in eq. \eqref{zetaTBA}.

Consider now adding an extra particle of type $r$ with rapidity $\theta_{0}$
to the condensate. The modified quantization condition to the condensate
reads
\bea
2\pi n_{j}=&\nonumber\\
&m_{s}L\sinh\theta_{j}+\delta_{rs}\left(\theta_{j}-\theta_{0}\right)+\sum_{k\neq j}\delta_{ss}(\theta_{j}-\theta_{k}), \nonumber\\
n_{j}\in\mathbb{Z}
\eea
In the continuum,
\bea
\frac{m_s}{2\pi}\cosh\theta=\tilde{\rho}\left(\theta\right)+\intop_{-B}^{B}d\theta^{\prime}\tilde{\rho}\left(\theta^{\prime}\right)\mathcal{K}_{ss}\left(\theta-\theta^{\prime}\right) \nonumber\\
-\frac{1}{2\pi L}\partial_{\theta}\delta_{rs}\left(\theta-\theta_{0}\right)
\eea
The equation for the density difference wrt. the ground state $D_{\theta_{0}}\left(\theta\right)=L\left[\tilde{\rho}\left(\theta\right)-\rho_s\left(\theta\right)\right]$
takes the form
\be
D_{\theta_{0}}\left(\theta\right)=\left(1-\mathcal{K}_{ss}\right)^{-1}\left[-\mathcal{K}_{rs}\left(\theta-\theta_{0}\right)\right]
\ee
with $\mathcal{K}_{rs}\left(\theta-\theta_{0}\right)=-\frac{1}{2\pi}\partial_{\theta}\delta_{rs}\left(\theta-\theta_{0}\right)$.
The energy difference due to the extra particle is then
\bea
&&E_{1}-E_{0}\equiv\epsilon_{r}\left(\theta_{0}\right)\nonumber\\
&&=m_{r}\cosh\theta_{0}-\mu Q_r+\intop_{-B}^{B}D_{\theta_{0}}\left(\theta\right)\left[m_{s}\cosh\theta-\mu\right]d\theta\nonumber\\
&&=m_{r}\cosh\theta_{0}-\mu Q_r-\intop_{-B}^{B}\mathcal{K}_{rs}\left(\theta-\theta_{0}\right)\epsilon_{s}\left(\theta\right)d\theta.
\eea
in agreement with eqs. \eqref{tba1}-\eqref{tba1v}. Here $Q_s=+1$, $Q_{\bar{s}}=-1$ and $Q_r=0$ otherwise.
When $r=s$, this yields the Fredholm equation for $\epsilon_{s}$.
For $r\neq s$, the spectral gap can be calculated straightforwardly
once $\epsilon_{s}$ is known. When an extra soliton of rapidity $\theta_{0}$
is added to the system, the net extra charge
\bea
Q_{1}-Q_{0}&=&1+\intop_{-B}^{B}D_{\theta_{0}}\left(\theta\right)d\theta\\
&=&1-\intop_{-B}^{B}\mathcal{K}_{ss}\left(\theta-\theta_{0}\right)\zeta\left(\theta\right)\equiv\zeta\left(\theta_{0}\right)
\eea
equals $\zeta\left(\theta_{0}\right)$. Hence the name ``dressed charge''.

\section{Numerical analysis for a single flavor}
\label{app_num}

The integral equations \eqref{tba1}-\eqref{zetaTBA} can be solved independently.
They are all Fredholm equations of the second kind, for which many numerical methods exist.
In fact TBA equations of this type can be solved well beyond machine precision (see e.g. \cite{Abbott:2020qnl}).
A simple yet efficient method is provided by the Nystrom method \cite{10.5555/1403886}. In case of an integer number of colors $N_c$, the kernel function $K(\theta)$ can be written in a closed form in terms of digamma functions:
\begin{widetext}
    \begin{eqnarray}
{\cal K}(\theta)=-\frac{1}{2\pi}\frac{d\delta(\theta)}{d\theta}
= \frac{1}{4\pi^2}
\sum_{n=0}^{\nu-2}&&\left[\psi\left(\frac12\left(1-\frac{i\theta}{\pi}+\frac{n}{\nu-1}\right)\right)
+ \psi\left(\frac12\left(1+\frac{i\theta}{\pi}+\frac{n}{\nu-1}\right)\right)
\right.  \nonumber \\
&-& \left. \psi\left(\frac{\pi+n\pi-i\theta(\nu-1)}{2\pi(\nu-1)}\right)
-\psi\left(\frac{\pi+n\pi+i\theta(\nu-1)}{2\pi(\nu-1)}\right)\right] 
    \end{eqnarray}
    \end{widetext}
    By a change of variables $\theta=B y$, $\theta^\prime=B y^\prime$, $\hat{\epsilon}(y)=\epsilon(B y)$, $\hat{G}(y,y^\prime)={\cal K}(B(y-y^\prime))$, we transform the integral equations to the interval $[-1,1]$.  For example, Eq. \eqref{tba1} takes the form
    \begin{equation}
        \hat{\epsilon}(y)+B\intop_{-1}^{1}dy^\prime \hat{G}(y,y^\prime)\hat{\epsilon}(y^\prime)=m_s \cosh(B y)-\mu ,\label{transfTBA}
    \end{equation}
    and the boundary condition becomes $\hat{\epsilon}(\pm 1)=0$.

    The integral in eq. \eqref{transfTBA} is then approximated by the Gauss-Legendre quadrature rule. This means that the integrand is to be evaluated at the roots $y_i$, $i\in{1,...,N}$ of the $N$th Legendre polynomial. To this end, we introduce $N$-component vectors $\hat{\epsilon}_i\equiv \hat{\epsilon}(y_i)$, $u_i=m_s\cosh(B y_i)-\mu$ and the matrix $\hat{G}_{ij}=\hat{G}(y_i,y_j)$. The components of the vector $\hat{\epsilon}(y_i)$ are obtained through the solution of the linear system
    \begin{equation}
        (1_{ij}+B w_j\hat{G}_{ij})\hat{\epsilon}_j=u_i.
    \end{equation}
    where $w_j$ are the quadrature weights.
    Finally, the solution is given by substituting back to eq. \eqref{transfTBA} and using the quadrature rule
    \begin{equation}
        \hat\epsilon(y)\approx m_s \cosh(B y)-\mu-B\sum_{i=1}^N w_i\hat{G}(y,y_i)\hat{\epsilon}_i
    \end{equation}
    This provides a robust way to reproduce the unknown function and its derivative.

\begin{figure}[t]
\begin{subfigure}[b]{0.475\textwidth}
\centering
\includegraphics[draft=false,width=\textwidth]{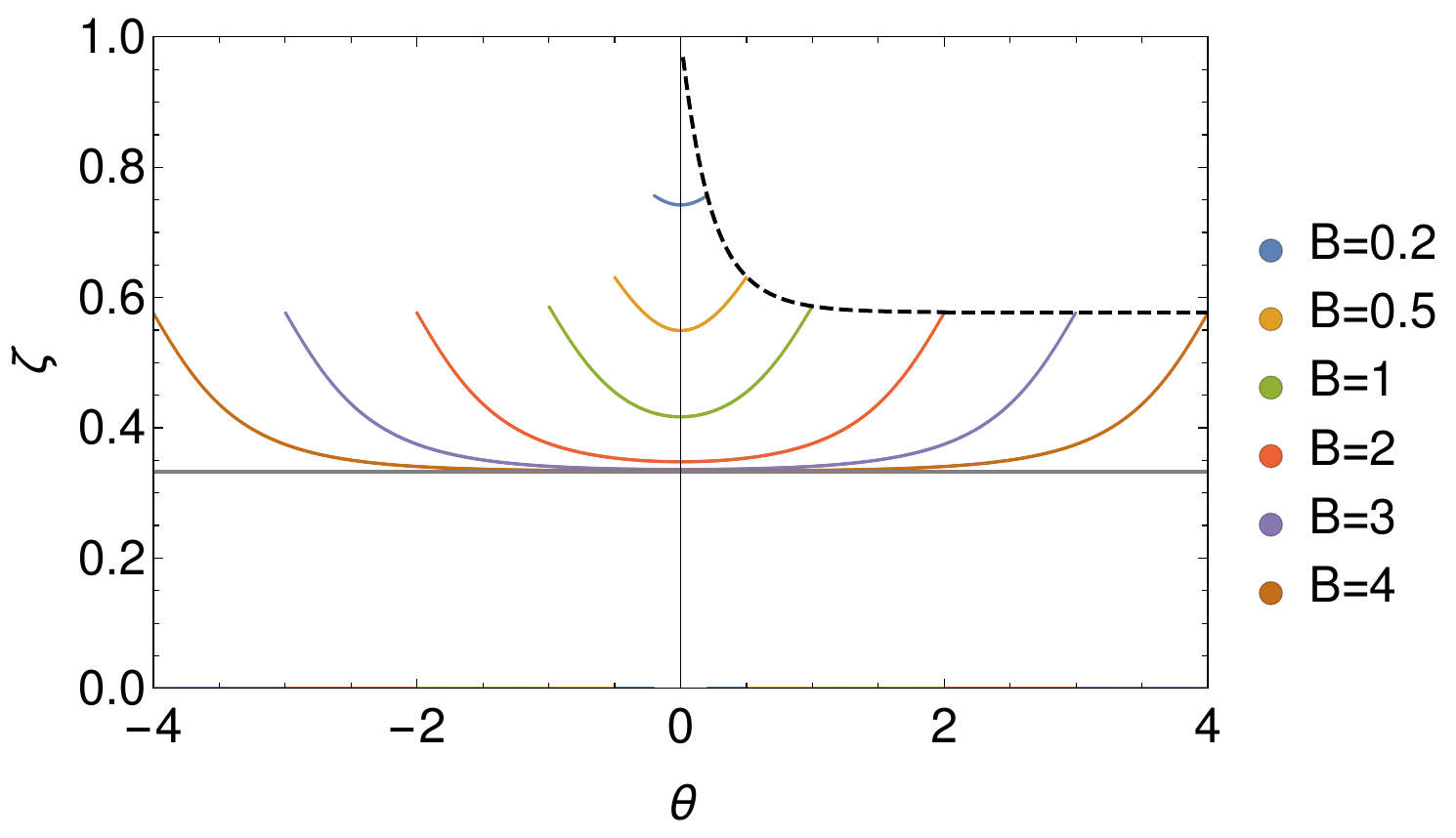}
\end{subfigure}
\hfill
\begin{subfigure}[b]{0.475\textwidth}
\centering
\includegraphics[draft=false,width=\textwidth]{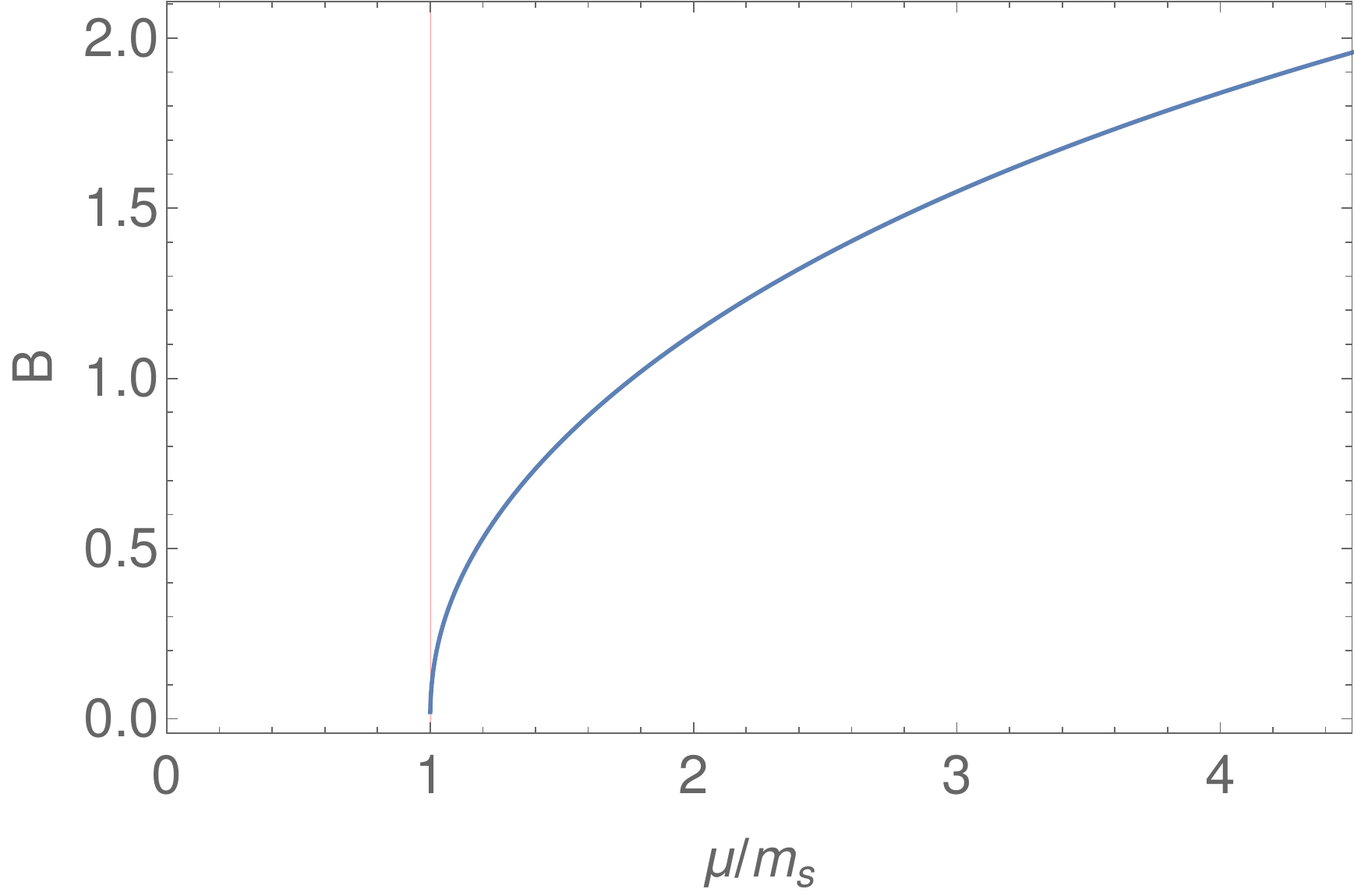}
\end{subfigure}
\caption{Numerical solution of the TBA equations (\eqref{tba1}-\eqref{zetaTBA})for $N_c=3$, $m_s=1$.
On the top panel, solutions for different $B$ values are shown. The $\zeta$-$B$ relation is drawn with a dashed curve ($\zeta=\sqrt{\tilde{K}}$ is the dressed charge). On the bottom, the relation $B(\mu)$ is shown.}\label{FigBdep}
\end{figure}

The results for $\zeta(\theta)$ are shown in Fig. (\ref{FigBdep}a). Note the plateau that develops for large values of $B$. Eq. \eqref{zetaTBA} has a constant bulk solution as $B\rightarrow\infty$, which can be obtained by a simple Fourier transform. Also plotted are the asymptotic values $\zeta_{bulk}=2/\nu$. This signals that the $B\rightarrow\infty$ limit has to be taken carefully as the behavior at the boundary is never reproduced from the bulk solution.

It is easy to obtain the (inverse) relation $\mu(B)$ from Eq. \ref{tba1}. A simple implementation of the secant method, connected to the Fredholm solver for $\epsilon(B,\mu)$, converges to the root $\epsilon(\mu)=0$ within a small number of iterations for each point. The result is depicted on Fig. (\ref{FigBdep}b). The $\zeta-\mu$ relation is plotted in Fig. (\ref{Spec}). There is a phase transition at $\mu=m_s(=1)$. The parameter $\zeta$ approaches $1$ in this limit. Increasing $\mu$, $\zeta$ quickly takes its asymptotic value, $\zeta_*=\sqrt{2/\nu}$.

We note that the relation $\mu(B)$ is available analytically through a large-$B$ expansion, which is provided in implicit form in \cite{Zamolodchikov:1995xk}. This was useful to check the accuracy of the numerical method.
The relation is given as
\bea
   && \mu=\frac{m_s \sqrt{\pi}\nu}{2}\frac{\Gamma\left(\frac{\nu}{2(\nu-1)}\right)}{\Gamma\left(\frac{1}{2(\nu-1)}\right)}q^{-\frac{\nu}{4(\nu-1)}}\times\nonumber\\
    && \frac{1}{1-\sum_{n=1}^\infty\frac{\nu}{1+(2n+1)(\nu-1)}q^n b_n w_n } ,\label{muexprq}
\eea
where the expansion is in terms of $q$,
\begin{eqnarray}
    q&=&e^{-4(B+\Delta)\left(1-\frac{1}{\nu}\right)};\\
    \Delta&=&\frac{1}{2}\log(\nu-1)-\frac{\nu}{2(\nu-1)}\log\nu .
\end{eqnarray}
The coefficients $b_n$ are given explicitly as
\begin{equation}
    b_n=\frac{(-1)^n}{n!(n-1)!}\frac{\Gamma\left(\frac{n}{\nu}\right)\Gamma\left(\frac{3}{2}+\frac{n(\nu-1)}{\nu}\right)}{\Gamma\left(-\frac{n}{\nu}\right)\Gamma\left(\frac{3}{2}-\frac{n(\nu-1)}{\nu}\right)},
\end{equation}
and $w_n\equiv w_n(q)$ are provided implicitly through the linear system
\begin{equation}
    w_n=\frac{1}{n}-\sum_{m=1}^\infty\frac{q^m}{m+n}b_m w_m.
\end{equation}

\begin{figure}[t]
\centering
\includegraphics[draft=false,width=0.475\textwidth]{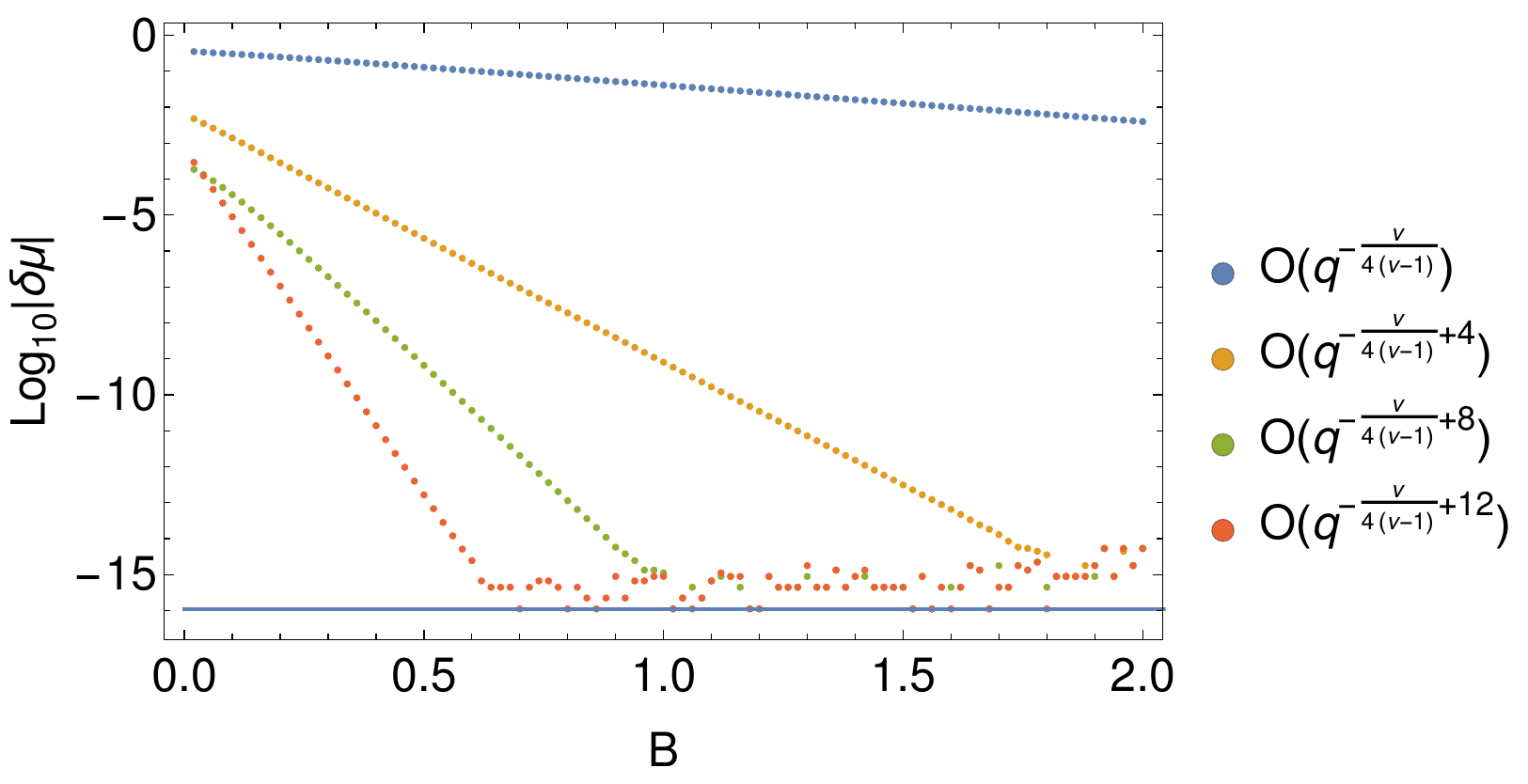}
\caption{Precision of various orders of the series expansion of eq. \eqref{muexprq}, as compared to TBA. The base-10 logarithm of \begin{tt}MachinePrecision\end{tt}
is shown with a blue line.}
\end{figure}

\section{Details of the perturbative calculation} \label{app_details}

Using Eq. \eqref{eq5v}, normalizing the vertex operators in finite volume with the appropriate $(2\pi L^{-1})$ factors
so that their normalization is consistent with eq. \eqref{twoptfnnormCFT},
\begin{eqnarray}
&& \chi(q,i\omega)_1 = \left(\frac{\tilde{m}}{4\pi}\right)^2 \frac{1}{N_cN_f\pi} \frac{\beta^2}{(4\pi)^2} \nonumber\\
&& \intop_{-\infty}^\infty d\tau_1\intop_{-\infty}^\infty d\tau_2\intop_{-\infty}^\infty\intop_{-\infty}^\infty dx_1 dx_2 \cdot \nonumber\\
&& \intop_{-\infty}^\infty\intop_{-\infty}^\infty dX dT e^{i q X+i\omega T}|z_1-z_2|^{-2\alpha}\cos(2k_0(x_1-x_2))\cdot \nonumber\\
&& \left(\frac{1}{Z-z_1}-\frac{1}{Z-z_2}-c.c.\right)\left(\frac{1}{z_2}-\frac{1}{z_1}-c.c.\right) \; , \label{leadordint}
\end{eqnarray}
where $Z=T-i X$, $z_{1,2}=\tau_{1,2}-i x_{1,2}$ and $\alpha$ was defined in eq. \eqref{alphadef}.

We first perform the $X$ and $T$ integration of eq. \eqref{leadordint}. Using the residue theorem, we obtain
\bea
&& \chi(q,i\omega)_1 = \left(\frac{\tilde{m}}{4\pi}\right)^2 \frac{1}{N_cN_f\pi}\frac{\beta^2}{(4\pi)^2}\nonumber\\
&& \intop_{-\infty}^\infty d\tau_2\intop_{-\infty}^\infty d\tau_1\intop_{-\infty}^\infty\intop_{-\infty}^\infty dx_1 dx_2 \nonumber\\
&& |z_1-z_2|^{-2\alpha}\cos(2k_0(x_1-x_2))\left(\frac{1}{z_2}-\frac{1}{z_1}-c.c.\right)\nonumber\\
&& \hskip .5in \times \frac{4\pi q }{q^2 +\omega^2}\left(e^{i q x_2 + i \omega \tau_2}-e^{i q x_1+ i\omega \tau_1}\right). \label{leadordint2}
\eea
We then change integration variables $x_1\rightarrow x+x_2$, $\tau_1\rightarrow \tau+\tau_2$ and integrate with respect to $x_2$ and $\tau_2$. This is again possible with the residue theorem. We arrive at
\bea
&& \chi(q,i\omega)_1=\left(\frac{\tilde{m}}{4\pi}\right)^2\frac{1}{N_cN_f\pi}\frac{\beta^2}{(4\pi)^2} \frac{4\pi q }{q^2+\omega^2}\frac{-16\pi q }{q^2+\omega^2}\nonumber\\
&& \intop_{-\infty}^\infty \rd\tau\intop_{-\infty}^\infty \rd x\nonumber\\
&& (x^2+\tau^2)^{-\alpha}\cos(2k_0 x)\sin^2\left(\frac{q x+\omega \tau }{2}\right) \; . \label{leadordint3}
\eea
This integral can be evaluated analytically, and yields an explicit formula for general $q,\omega$, leading to the result reported in Eq. \eqref{PertResult}.

%

\end{document}